\documentclass[structabstract]{aa}  
\usepackage{graphicx}
\usepackage{natbib}
\usepackage{txfonts}
\begin{document}
   \title{Warm dusty discs: Exploring the A star 24$\mu$m debris population}

   \author{R. Smith
          \inst{1,2}
          \and
          M.C. Wyatt \inst{2}
          }

   \offprints{R. Smith}

   \institute{Astrophysics Group, Keele University, Keele,
     Staffordshire, ST5 5BG \\
              \email{rs@astro.keele.ac.uk}
             \and
              Institute of Astronomy, University of Cambridge,
	      Madingley Road, Cambridge, CB3 0AH }

   \date{Revised 8th July 2009}

 
  \abstract
   {}
   {Studies of the debris disc phenomenon have shown that most systems
  are analogous to the Edgeworth-Kuiper Belt (EKB). In this study we
  aim to determine how many of the IRAS 25$\mu$m excesses towards A
  stars, which may be indicative of asteroid belt analogues, are real,
  and investigate where the dust must lie and so build up a picture of
  what these systems are like.}
   {We observe using ground-based mid-infrared imaging with TIMMI2,
     VISIR, Michelle and TReCS a sample of A and B-type main sequence
     stars previously reported as having mid-infrared excess. We
     combine modelling of the emission spectrum from multi-wavelength
     photometry with a modelling technique designed to constrain the
     radial extent of emission in mid-infrared imaging to constrain
     the possible location of the debris.} 
   { We independently confirm the presence of warm dust around three of the
  candidates: HD 3003, HD 80950 and $\eta$ Tel.  For the binary HD3003
  a stability analysis indicates the dust is either circumstellar and
  lying at $\sim$ 4 AU with the binary orbiting at $>$ 14AU, or the
  dust lies in an unstable location; there is tentative evidence for
  temporal evolution of its excess emission on a $\sim$20 year
  timescale.  For 7 of the targets we present quantitative limits on
  the location of dust around the star based on the unresolved
  imaging. We demonstrate that the disc around HD71155 must 
  have multiple spatially distinct components at 2 and 60AU.   
  We model the limits of current instrumentation to resolve debris
  disc emission and show that most of the known A star debris discs
  which could be readily resolved at 18$\mu$m on 8m instruments have
  been resolved, but identify several that could be resolved with deep
  ($>$ 8 hours total) integrations (such as HD19356, HD139006 and
  HD102647). }
   {Limits from unresolved imaging can help
  distinguish between competing models of the disc emission, but
  resolved imaging is key to an unambiguous determination of the disc
  location. Modelling of the detection limits for extended emission can be
  useful for targeting future observational campaigns towards sources
  most likely to be resolved. MIRI on the JWST will be able to resolve the
  majority of the known A star debris disc population.  METIS on the
  E-ELT will provide the opportunity to explore the hot disc
  population more thoroughly by detecting extended emission down to where
  calibration accuracy limits disc detection through photometry alone,
  reaching levels below 1 zodi for stars within 10pc. }


   \maketitle
%

\section{Introduction}
\label{s:intro}

Analysis of the IRAS database over the last 20 years has shown that
over 300 main sequence stars have dust discs around
them.  This material is thought to be the debris left over at the end
of the planet formation process \citep[e.g.][]{mannings}. 
The spectral energy
distribution (SED) of this excess in the best studied cases 
(e.g., Vega, $\beta$ Pictoris, Fomalhaut, $\epsilon$ Eridani)
peaks longward of 60$\mu$m implying that this dust is cool ($<$80K), and so
resides in Edgeworth-Kuiper belt (EKB)-like regions in the systems. 
The EKB-like location and analogy is confirmed in the majority
  of cases where
these discs have been resolved (in thermal emission,
  e.g. \citealt{holland, greaves}; in scattered light imaging, e.g.
\citealt{kalas07, schneider, schneider09}), since
the dust is shown to lie $>$40AU from the stars, and 
its short lifetime means that it must be continually replenished by
the collisional destruction of km-sized planetesimals
\citep{wyattdent}. The inner 40AU radius hole is thus thought to arise from
clearing by an unseen planetary system, the existence of which is
supported by the presence of clumps and asymmetries seen in the
structure of the dust rings \citep[e.g.][]{wyatt99, greaves98, augereau}. 

On closer examination the inner regions of known cool debris disc
systems are much more complex than simply 'dust-free inner holes'. 
$\beta$ Pictoris has (a
relatively small amount of) resolved dust in this inner region 
\citep{lagage, telesco, boccaletti}, thought to be there because this is
a young (12Myr, \citealt{zuckermanEA}) transitional system in which
these regions have yet to be fully cleared by planet formation
processes. \citet{absil} have recently presented
interferometric data showing Vega (thought to be around 380-500Myr old,
\citealt{peterson}) is likely to possess extended dust
emission within 8AU, with evidence for a similar warm dust propulation
  around Fomalhaut \citep{absil09}, and \citet{difolco} have also recently
presented evidence for hot dust around the 10Gyr old $\tau$ Ceti.
Multiple-component discs mirror our own Solar System, with our debris
disc concentrated in the asteroid belt and EKB.  It
is possible that more sources with known cold discs have dust in the
inner system, but the difficulty of separating hot dust emission from
the stellar photosphere often limits detection. 
\citet{chen09} presented a sample of 11 debris systems believed to have
multiple dust belts, although confirmation of the multiple components
via resolved imaging is required for several of these stars.

The presence of excess in the mid-infrared range is of particular
interest.  The temperature of dust emitting at 24$\mu$m suggests that
it should lie close to the star, in regions of a few to a few tens of
AU.  These are regions in which we might expect planets to reside (see
above), and so the origin of the emission and possible links with any
cold debris in the system must be explored.  Surveys suggest that
24$\mu$m excess may be common, with around half of main sequence stars
that exhibit excess mid-infrared emission in the IRAS database
\citep{mannings} having  an excess at 25$\mu$m only \citep{zuckerman}.
However, excesses taken from the IRAS database cannot be used at face
value.  \citet{song02}, who searched the IRAS database for excess
emission towards M-type stars, noted that when searching a large 
number of stars for excesses close to the detection threshold, a
number of false positives must be expected due to noise.  There have
also been a few instances in which the IRAS excess has been shown to 
be attributed to background objects that fall within the relatively
large IRAS beams ($>$30''; such objects range from highly reddened
carbon stars or Class II YSO's \citealt{lisse}, to distant galaxies
\citealt{sheret}). Another possible source of mid-infrared excess
emission is reflection nebulosity \citep{kalas}.  Indeed it is now
routine for papers discussing the excess sources found by IRAS to
address the possibility that some of these are bogus debris discs
\citep{moor, rhee}. 

Recent statistical studies using Spitzer data have revealed a large
sample of A stars with excess emission at 24 and/or 70$\mu$m
\citep{rieke,su}.  Spitzer observations benefit from smaller beamsizes
and higher resolution compared to IRAS, improving the reliability of excess
measurements.  The 24 and 70$\mu$m excesses around A stars have a
wide variation in levels amongst systems of similar ages, but overall
there is a decrease in the upper envelope of excess inversely 
proportional to
time \citep{rieke,su}.  These features can be interpreted in terms of
a steady-state evolution of belt-like planetesimal discs in a
collisional cascade, where the fall-off with time is due to the
collisional grinding away of material and the variation in excess
levels between systems of similar ages can be reproduced by variations
in initial disc mass and planetesimal belt radii \citep{wyattsmith07}.   
Although recent work by \citet{Currie} has shown that excess emission at
24$\mu$m around A stars increases from 5--10 Myr and peaks around 10--15
Myr before declining with age, which is not predicted by the
steady-state model, this can be explained by the delayed formation of
Pluto-sized bodies in the disc \citep{kenyon04II}. It is only when
Pluto-sized bodies are formed that the orbits of planetesimals are
stirred to high relative velocities and the steady-state collisional
cascade can begin.  

However this model does not account for the possibility of multiple
belts (see above).  There have also been other models proposed to
explain the variance in excess levels between similarly aged stars and
the variety in the SED slopes of those stars with excess at both 24
and 70$\mu$m in \citet{rieke}, which can be interpreted as evidence
for stochastic evolution \citep{rhee}.  A collision similar to the
Earth-Moon forming massive collision has been proposed as the
explanation for the large excess and spectral features observed around
HD172555 \citep{lisse09}. Similarly short-lived dust production origins have
been proposed for dust found close to the central star around several
Sun-like stars. For these systems a steady-state collisional cascade
production  from a spatially coincident planetesimal belt cannot
explain the levels of excess \citep{wyattsmith06, lohne}.  For debris discs
with a cold dust population in addition to hot dust emitting at
24$\mu$m the outer planetesimal belt could be feeding the hot dust
population (as has been proposed for $\eta$ Corvi, \citealt{wyattsmith06,
  smithhot, smithMIDI}).  However, the mechanism that might be
transporting the dust from cold outer regions to the hot inner
locations observed at 24$\mu$m is as yet unclear (possibilities
include dynamical scattering by a migrating planet, see
e.g. \citealt{gomes, booth}).   

To tackle these issues regarding the origin of the 24$\mu$m dust
emission, and in particular explore what its presence may reveal
about planet formation and as yet undetected planetary populations, 
this paper looks at the constraints on the true dust distribution 
around a sample of A stars with 24$\mu$m excess.  
This can be assessed indirectly from SED fitting to multi-wavelength
infrared photometry, but uncertainties arising from degeneracies
in dust model fitting and the possibility of multiple temperatures of
dust mean that determining radial location from SED fitting alone is
challenging (see discussions of individual sources in section
4). Resolved imaging provides more direct constraints on the dust
location.

This paper is structured as follows; In \S \ref{s:sample} the sample
selection is described. In \S \ref{s:obsext} we describe the various
observational and analysis techniques employed for the observations,
with the results and discussion of individual sources 
presented in \S \ref{s:res}.  An extension modelling technique is used
to explore which of the A star discs in the literature
may be fruitful subjects of future imaging in \S \ref{s:pred}. 
The implications of these and the observational results
are discussed in \S \ref{s:dis}. Conclusions are in \S \ref{s:conc}.

%

\section{The Sample}
\label{s:sample}

The sample consists of A and B stars with IRAS published detections
of excess emission at 12 and/or 25 $\mu$m.\footnote{The sample stars
  are listed in the Debris Disc Database at
http://www.roe.ac.uk/ukatc/research/topics/dust.}
A first-cut was applied to the list of all published detections to
produce a final sample of 11 candidates (Table \ref{sample}). This
first-cut consisted of the analysis outlined below to determine if the
excess identified by IRAS was likely to be real. 

For each star J, H, and K band fluxes were obtained from 2MASS
\citep{2mass} and V and B magnitudes from Tycho2
\citep{tycho}.  The photospheric emission was then 
determined by adopting a Kurucz model \citep{kurucz} for the
appropriate spectral type (as listed in the Michigan Spectral
Catalogues or SIMBAD) scaled to the K band flux.  IRAS fluxes were
extracted using SCANPI  (the Scan Processing and 
Integration tool)\footnote{http://scanpi.ipac.caltech.edu:9000/}.
The expected stellar flux was multiplied by the colour-correction
factor (at the levels described in the IRAS Explanatory
Supplement\footnote{The IRAS Explanatory Supplement is available at
  http://irsa.ipac.caltech.edu/IRASdocs/exp.sup/}) before subtraction
from the IRAS flux to determine the excess.  The proximity
of the IRAS sources to the stars was also checked given the quoted
uncertainty error ellipse, since some surveys allowed excess sources
to be up to 60 \arcsec offset and have since been shown to not be related
\citep{sylvester}.  

\begin{table*}
\caption{The Sample}
\label{sample}
\centering                          
\begin{tabular}{|c|c|c|c|l|l|} \hline Star name & Stellar type & Age &
Distance$^a$ &  \multicolumn{2}{|c|}{IRAS fluxes (mJy)$^b$}
\\ 
 &  & Myr & pc & 12$\mu$m & 25$\mu$m   \\ \hline
HD 3003 & A0V & 50$^c$ & 47 & 148 + 16 ($\pm$20) & 34 + 241 ($\pm$20)
\\  
HD 23281 & A6V & 626$^d$ & 43 & 251 + 36 ($\pm$20) & 59 + 107
($\pm$28)  \\  
HD 23432 & B8V & 100$^e$ & 119 & 127 + 256 ($\pm$43) & 29 + 1159
($\pm$42)  \\ 
HD 31295 & A0V & 100$^e$ & 37 & 336 + 238 ($\pm$65) & 78 - 110
($\pm$121)  \\ 
HD 38206 & A0V & 9$^f$ & 69 & 130 + 7 ($\pm$23) & 30 + 67 ($\pm$21)
\\ 
$\lambda$ Gem$^g$ & A3V & 560$^f$ & 29 & 1166 + 214 ($\pm$28) & 271 +
161 ($\pm$52) \\ 
HD 71155 & A0V & 169$^c$ & 38 & 675 + 162 ($\pm$31) & 157 + 249
($\pm$41)  \\ 
HD 75416 & B8V & 5$^f$ & 97 & 134 + 86 ($\pm$22) & 31 + 86 ($\pm$38)
\\ 
HD 80950 & A0V & 80$^f$ & 81 & 120 - 5 ($\pm$26) & 28 + 101 ($\pm$17)
\\ 
HD 141795 & Am & 450$^e$ & 22 & 1154 + 67 ($\pm$25) & 269 + 141
($\pm$28)  \\ 
$\eta$ Tel$^i$ & A0V & $12^h$ & 48 & 263 + 138 ($\pm$34) & 61 + 394
($\pm$22)  \\ 
\hline
\end{tabular}

Notes: $^a$=Distance from parallax in Hipparcos; $^b$=Fluxes are shown as star +
 excess ($\pm$error);  $^c$=Age taken from \citet{song01}; $^d$=Age taken
 from \citet{kunzli}; $^e$ = Age determined from
 theoretical evolutionary tracks of \citet{song01}, see discussion
 in text; $^f$=Age taken from \citet{rieke}; $^g$ HD 56537; 
 $^h$= =Age from $\beta$ Pictoris association membership,
 \citet{zuckermanEA}; $^i$ HD 181296.  For those sources without ages
 listed in the literature (HD 23432, HD 31295 and HD 141795) the $\beta$ and
$c_1$ magnitudes as listed in \citet{hauck} were used to
determine $T_{\rm{eff}}$ and $\log(g)$ from the grids of \citet{moon}.
Then the theoretical evolutionary tracks of \citet{song01} were used
to determine the approximate ages for these stars.

\end{table*}

%

\section{Observations and Data Reduction}
\label{s:obsext}

\subsection{Observations}
\label{s:obs}

Observations were performed using: TIMMI2 on the ESO 3.6m telescope at
La Silla (proposals 71.C-0312, 72.C-0041 and 74.C-0070); VISIR on the
ESO VLT (proposal 076.C-0305); and Michelle and TReCS on the twin
telescopes of Gemini (GN-2005B-Q-15 and GS-2005B-Q-67). All
observations used chop-nod pattern to remove 
sky and telescope emission.  A chop of 10\arcsec in the North-South
direction with a perpendicular 10\arcsec nod was used for the ESO
observations.  The Gemini observations used a 15\arcsec chop and
parallel nod (also of 15\arcsec) at 30$^\circ$ East of
North. Observations of $\lambda$ 
Gem were performed with a chop at 268$^\circ$ to ensure the image of
the binary companion would fall on the array. 

This chop-nod pattern means that a simple co-addition of the data
produces an image with two positive and two negative images of the
source for the ESO observations, and one positive and two negative
images at half the intensity level for the Gemini observations.  A
dark current offset is determined from median values for 
each row and column of the image (excluding pixels on which the source
image fell) and subtracted from the final frame.  Pixels showing high
levels of variation throughout the observation (10 times the average)
were masked off.  Pixels showing very high or low gain (determined by
comparing average sky emission detected across the image to that
detected in each individual pixel detection) were also masked. In
total an average of around 7\% of 
pixels were removed in the TIMMI2 observations, and around 4\% of
pixels in the MICHELLE, TReCS and VISIR observations.  Calibration
observations of standard stars within a few degrees of the science
object were taken immediately before and after science observations. 
The standards were chosen from the list of K and M giants identified by
\citet{cohen}.   In addition to photometric calibration, these
standards were used to characterise the PSF and used for comparison
with the science sources to detect any extension (see
section \ref{s:extlim}).

\subsection{Photometry and Background/Companion Objects}
\label{s:phot}

The multiple images resulting from the chop-nod pattern  were
co-added to get a final image by first determining the centroid of each
of the individual images. Photometry was then performed using a
1\arcsec \, radius aperture for the TIMMI2 images and a 0\farcs5
radius aperture for the VISIR and MICHELLE images.  These sizes were
chosen to just exceed the full-width at half-maximum (FWHM) found for
each instrument (average FWHM: 0\farcs80 $\pm$ 0\farcs12 in the N
band, and 1\farcs34 $\pm$ 0\farcs10 at Q for TIMMI2; 0\farcs465 $\pm$
0\farcs161 at N on VISIR, and in the Q band 0\farcs597 $\pm$
0\farcs166; for Michelle 0\farcs557
$\pm$ 0\farcs107 in N and 0\farcs579 $\pm$ 0\farcs101 in Q; and for
the N band observations with TReCS 0\farcs475 $\pm$ 0\farcs054).  Note
that the filters used in these observations were narrow band and so 
no colour-correction was applied.  Residual statistical image noise
was calculated using an annulus centred on the star with inner radius
matching the outer radius of the aperture used for the photometry, and 
outer radius of twice the inner radius (so 2\arcsec \, for TIMMI2 and
1\arcsec \, for VISIR and MICHELLE).  Typical levels for statistical
noise at the 1 $\sigma$ level in a half hour observation were
44mJy total in the 1\farcs0  radius aperture of TIMMI2, 4 mJy and
12mJy for the 0\farcs5 aperture of VISIR in N and Q respectively, 
6mJy in the 0\farcs5 aperture of MICHELLE and 3 mJy in the 0\farcs5
aperture of TReCS. Calibration uncertainty was determined from
variation in standard star photometry, and was added in quadrature to
statistical uncertainty to give the total error on the photometry as
listed in Table \ref{observations}.

Smaller apertures were used to search for background sources and to
place limits on undetected sources.  The aperture radius was
  determined through examination of the standard star images.
  Circular apertures of increasing radius were centered on the
  standard star images and the radius giving the highest
  signal-to-noise (statistical noise only, as determined in the annuli
  listed above) was recorded.  For each instrument and observing
  wavelength the median optimal radius for maximising signal-to-noise
  on the standard stars was chosen to optimise point source
  detection.  The apertures
had radii of: 0\farcs8 for TIMMI2 observations; 0\farcs4 for MICHELLE
and for TReCS; and 0\farcs32 and 0\farcs35 for the N and Q filters for
VISIR. Apertures were systematically centred on each pixel of each
array to search for $>3\sigma$ detections; where none were found the
limits on any background object were based on the 3$\sigma$
uncertainty in the aperture plus calibration uncertainty.  For the
non-photometric nights, limits were 
based on calibration to the IRAS flux of the object.  The upper limits
to background sources are listed in Table \ref{observations}. These
have been translated into a limit on the spectral type of any
companion source.  These spectral limits assume any companion is a
main-sequence star at the same distance as the source, and are given
as the hottest star that does not exceed the point source limits found
in the imaging.  

\subsection{Extension testing and limits on disc size}
\label{s:extlim}

Evidence for extended emission was checked for all science
targets. 
The source's surface brightness profile was determined by calculating
the average surface brightness in a series of annuli centred on the
source of 2 pixel thickness by increasing inner radius from 0 to
3\arcsec, and this was compared to profiles of the standard targets.
Finally the images of the point-like standard stars scaled to the peak
of the science images were subtracted from the science images and the
residuals checked for consistency with noise levels measured on the
pre-subtraction image.   A range of regions optimised for
  different disc geometries were tested for evidence of residual
  flux indicating spatially extended emission.  These optimal regions
  were determined from extensive modelling work and are outlined in
  detail in section 4 of \citet{smithhot}.  

\begin{table*}
\caption{The Observations}
\label{observations}
\centering                          
\begin{tabular}{*{11}{|c}|} \hline  &
  \multicolumn{4}{|c|}{Observation} & Predicted & 
  \multicolumn{5}{|c|}{Results$^{a}$} \\ Star &  &
  & Integration & &  stellar  & Flux, & Total & Stats. &
  \multicolumn{2}{|c|}{Background limit}  \\ name & Filter &
  $\lambda$, $\mu$m & time, s & Instrument & 
  flux, mJy$^\ast$ & mJy & err., mJy & err., mJy & mJy$^b$ &
  Spc. Type$^c$ \\ \hline  
HD 3003 & N9.8 & 9.56 & 1800 & TIMMI2 & 426 & 481 & 87 & 17 & $\le$ 46
& M3.5V \\
     & N12.9 & 12.21 & 3600 & TIMMI2 & 262 & 436 & 64 & 10 & $\le$ 27
& \\
     & Q2 & 18.75 & 3100 & TIMMI2 & 112 & 206 & 140 & 44 & $\le$ 148 &
\\
     & Si-5 & 11.66 & 3400 & TReCS & 288 & 375 & 34 & 4 & $\le$ 10 & \\
     & Q2 & 18.72 & 2400 & VISIR & 113 & 244 & 27 & 9 & $\le$ 21 & \\ \hline
HD 23281 & N12.9 & 12.21 & 3200 & TIMMI2 & 243 & 339 & 57 & 10 & $\le$
27 & M5V \\
      & SiC & 11.85 & 3600 & VISIR & 258 & 252 & 13 & 2 & $\le$ 4  & \\
      & Q2 & 18.72 & 3800 & VISIR & 104 & 129 & 9 & 6 & $\le$ 13 & \\ \hline
HD 23432 & N7.9 & 7.77 & 1800 & TIMMI2 & 299 & 452 & 192 & 192 & $\le$
351 & A6V\\
      & N11.9 & 11.59 & 1800 & TIMMI2 & 136 & 141 & 62 & 14 & $\le$ 44
& \\ \hline
HD 31295 & N1 & 8.60 & 1800 & TIMMI2 & 648 & 851 & 77 & 13 & $\le$ 33
& M3V \\ \hline
HD 38206 & N9.8 & 9.56 & 1800 & TIMMI2 & 204 & 196 & 48 & 16 & $\le$
44 & G3.5V\\ \hline
$\lambda$ Gem & N1 & 8.60 & 1800 & TIMMI2 & 2244 & 1997 & 167 & 22 &
$\le$ 56 & M5.5V \\
Chop PA & SiC & 11.85 & 540 & VISIR & 1195 & 701 & 9 & 4 & $\le$ 7 & \\
268$^\circ$ & Q2 & 18.72 & 1880 & VISIR & 484 & 590 & 71 & 6 & $\le$
14 & \\
Binary & N1 & 8.60 & 1800 & TIMMI2 & 101 & 68 & 8 & & - & \\
Chop PA & SiC & 11.85 & 540 & VISIR & 54 & 46 & 4 &  & - & \\
268$^\circ$ & Q2 & 18.72 & 1800 & VISIR & 22 & 31 & 17 &  & - & \\ \hline
HD 71155 & N2 & 10.68 & 3060 & TIMMI2 & 850 & 1278 & 215 & 36 & $\le$
98 & K4.5V \\
      & Si-5 & 11.60 & 1410 & MICHELLE & 722 & 1050 & 115 & 12 & $\le$
32 & \\
      & Qa & 18.50 & 2100 & MICHELLE & 286 & 398 & 99 & 13 & $\le$ 38
& \\
      & Q2 & 18.72 & 3600 & VISIR & 280 & 380 & 36 & 7 & $\le$ 16 & \\ \hline  
HD 75416 & N11.9 & 11.59 & 1800 & TIMMI2 & 144 & 204 & 88 & 15 & $\le$
48 & F0V \\ \hline
HD 80950 & N11.9 & 11.59 & 3600 & TIMMI2 & 129 & 175 & 16 & 10 & $\le$
25 & K4.5V \\
      & SiC & 11.85 & 1800 & VISIR & 124 & 120 & 23 & 3 & $\le$ 7 & \\
      & Q2 & 18.72 & 3760 & VISIR & 50 & 119 & 11 & 9 & $\le$ 20 & \\ \hline
HD 141795 & N12.9 & 12.21 & 1200 & TIMMI2 & 1115 & 1138 & 119 & 19 &
$\le$ 49 & M4.5V \\
       & Qa & 18.50 & 1900 & MICHELLE & 491 & 511 & 39 & 3 & $\le$ 8 &
\\ \hline
$\eta$ Tel & N12.9 & 12.21 & 2500 & TIMMI2 & 254 & 351 & 23 & 10 &
$\le$ 25 & K2V \\
       & Q2 & 18.75 & 1800 & TIMMI2 & 109 & 391 & 182 & 43 & $\le$ 136
& \\
\hline
\end{tabular}

$^\ast$ The expected photospheric emission is determined by a Kurucz model
profile appropriate to the spectral type of the star and scaled to the
K band 2MASS magnitude as outlined in section 2 unless otherwise
stated in the individual source description. Errors are 1$\sigma$.  M
band TIMMI2 observations were largely non-photometric and primarily
used to improve pointing accuracy and thus are not listed in this
table.  
Notes: $^a$ Errors are total errors (inclusive of calibration
uncertainty and image noise). $^b$ Limits are 3$\sigma$ upper limit to
undetected object including calibration errors, or scaled to IRAS
fluxes when conditions were non-photospheric.  These limits are valid
to within 28\arcsec of the detected source for TIMMI2 observations,
12\farcs6 for MICHELLE, 8\farcs1 for TReCS, and 11\farcs4
of the source for VISIR observations. $^c$ The hottest spectral type
of main-sequence star equidistant with the target that would have
remained undetected within our field of view.  See section 3.2 for
details of this limit.  
\end{table*}

\begin{table*}
\caption{The Results}
\label{results}
\begin{center}                          
\begin{tabular}{*{8}{|c}|} \hline Star name &
 \multicolumn{3}{|c|}{Fit as dust disc} & 
\multicolumn{2}{|c|}{Limit on extension} & 
$f_{IR} = L_{\rm{dust}}/L{\ast}$ & $f_{\rm{max}}^a$ \\  & Temp, K &
 AU & \arcsec  & AU & $D_{\rm{grain}}, \mu$m$^b$ & $\times 10^{-5}$ &
 $\times 10^{-5}$  \\ \hline 
\multicolumn{8}{|l|}{Photometric confirmation of excess} \\ 
HD 3003 & 265 & 4.00 & 0.086 & $<6.5^{+2.1}_{-0.9}$ & 0.55 & 20.1 & 0.053 \\
HD 80950 & 180 & 13.6 & 0.17 & $<24.5^{+7.5}_{-3.5}$ & 0.01 & 9.62 & 0.544 \\
$\eta$ Tel$^c$ &  & 3.90 & 0.081 & $<6$ & & 15.7 & 0.210 \\ 
       &  & 21--26 & 0.44--0.54 & - & & 13.9 & 12.52 \\ \hline
\multicolumn{8}{|l|}{Constraints on radial extent of disc} \\ 
HD 71155 & 500 & 1.98 & 0.052 & $<8.2^{+5.7}_{-0.5}$ & - & 8.95 & 0.0034 \\
      & 90 & 61.03 & 1.59 & - & - & 1.92 & 7.60 \\
$\lambda$ Gem$^d$ & 420 & 2.16 & 0.07 & $<6.1^{+0.2}_{-0.3} $ & 0.012 &
5.64 & 0.002  \\ 
HD 23281 & 210 & 5.36 & 0.12 & $<9.5^{+4.2}_{-0.7}$ & 0.68 & 3.82 & 0.024  \\
HD 75416 & 250 & 11.1 & 0.11 & $<55.2^{+1.5}_{-1.5}$ & - & 7.98 & 2.44 \\
HD 141795 & 250 & 4.64 & 0.22 & $<6.1^{+0.3}_{-0.4} $ & 0.73 & 4.43 & 0.008 \\ \hline
\multicolumn{8}{|l|}{Unconstrained by these observations} \\ 
HD 31295 & 80 & 52.6 & 1.4 & - & - & 5.85 & 9.216 \\
HD 38206 & 90 & 48.4 & 0.70 & - & - & 15.1 & 84.96 \\ \hline
\end{tabular} \\
\end{center}
Note that the objects with no extension limits have too low a
fractional excess for the extension to have been detected in the
images regardless of size. Estimates of radius are based on blackbody
fits and could be up to three times larger if the grains responsible
for the emission are small. Limits on extension shown here are for a
narrow face-on discs. Errors on this limit arise from 3 $\sigma$
photometric errors and errors on the determination of the photospheric
emission at the observed wavelength.   These sources are combined to
give 3$\sigma$ uncertainties on $F_{\rm{disc}}$, which in turn give a
range of radii at which a disc could have been detected as extended emission. 
Horizontal lines indicate division into photometrically
confirmed debris discs, and sources
for which our results provide constraints on the discs (sections 4.1 and
4.2 respectively). \\
Notes: $^a$ See section \ref{s:dis} for details of this limit; $^b$
the minimum grain diameter (for the composition assumed in section
3.3) such that grains at the maximum radius
given by the extension testing limits do not exceed the temperature of
the blackbody fit to the excess emission (where limits on the disc radius
are large compared to the blackbody radius no limit can be placed on the
minimum grain size, as even the smallest grains cannot reach the
fitted temperature at the extension limit radius); $^c$
These disc locations are taken from \citet{smitheta}; $^d$ binary detected. 
\end{table*}

\begin{figure*}
\begin{minipage}{8cm}
\includegraphics[width=8cm]{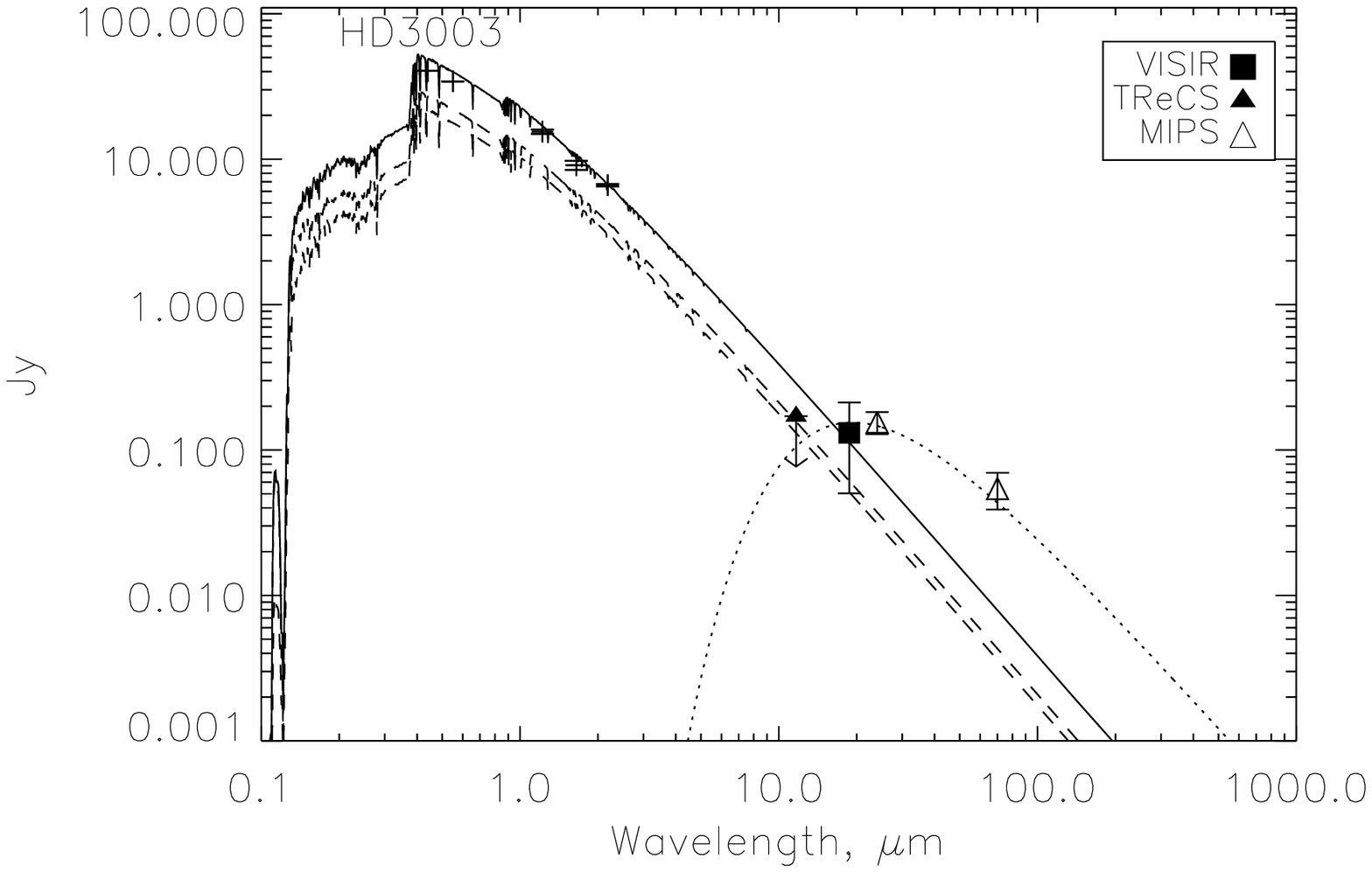}
\end{minipage}
\hspace{1cm}
\begin{minipage}{8cm}
\includegraphics[width=8cm]{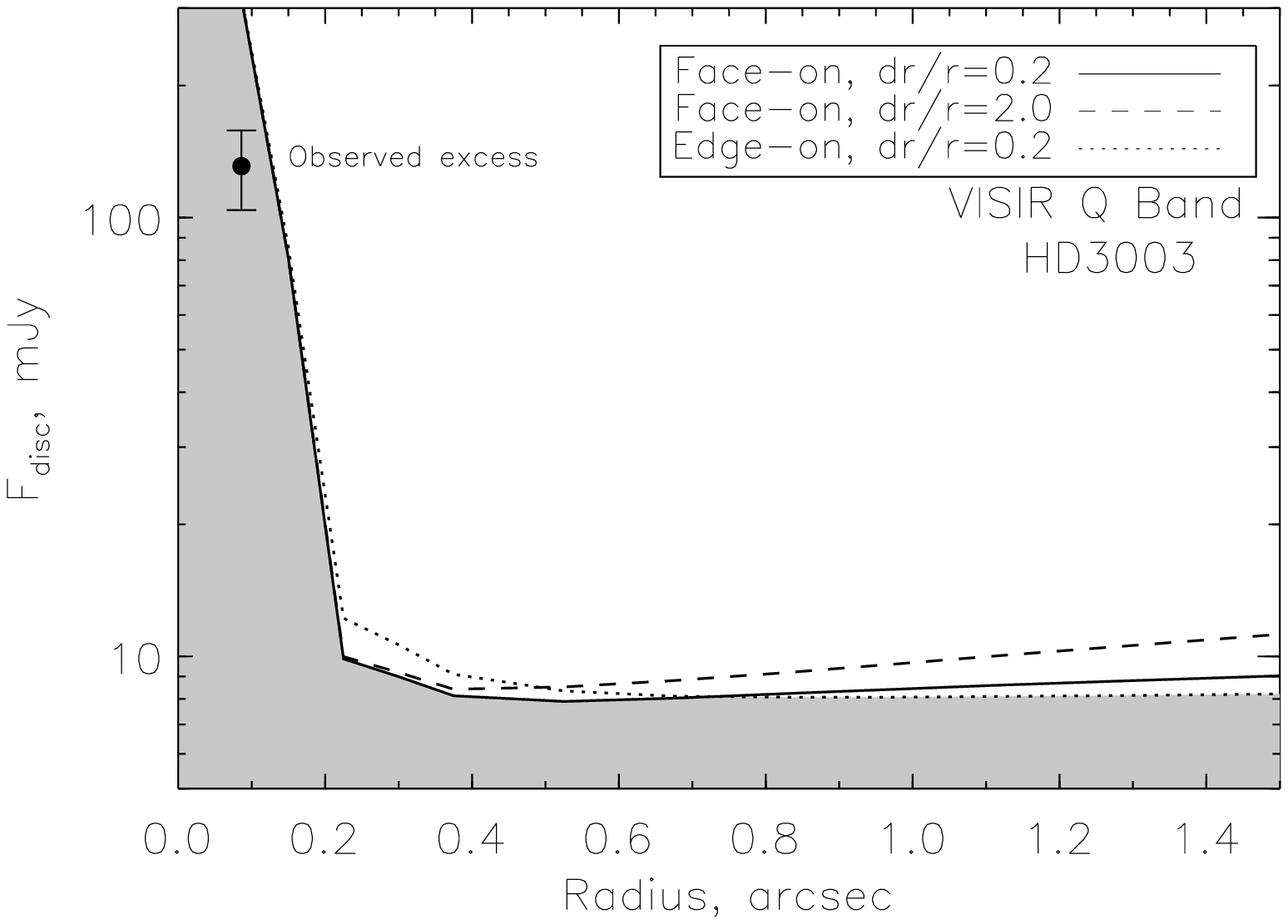}
\end{minipage}
\\ 
\begin{minipage}{8cm}
\includegraphics[width=8cm]{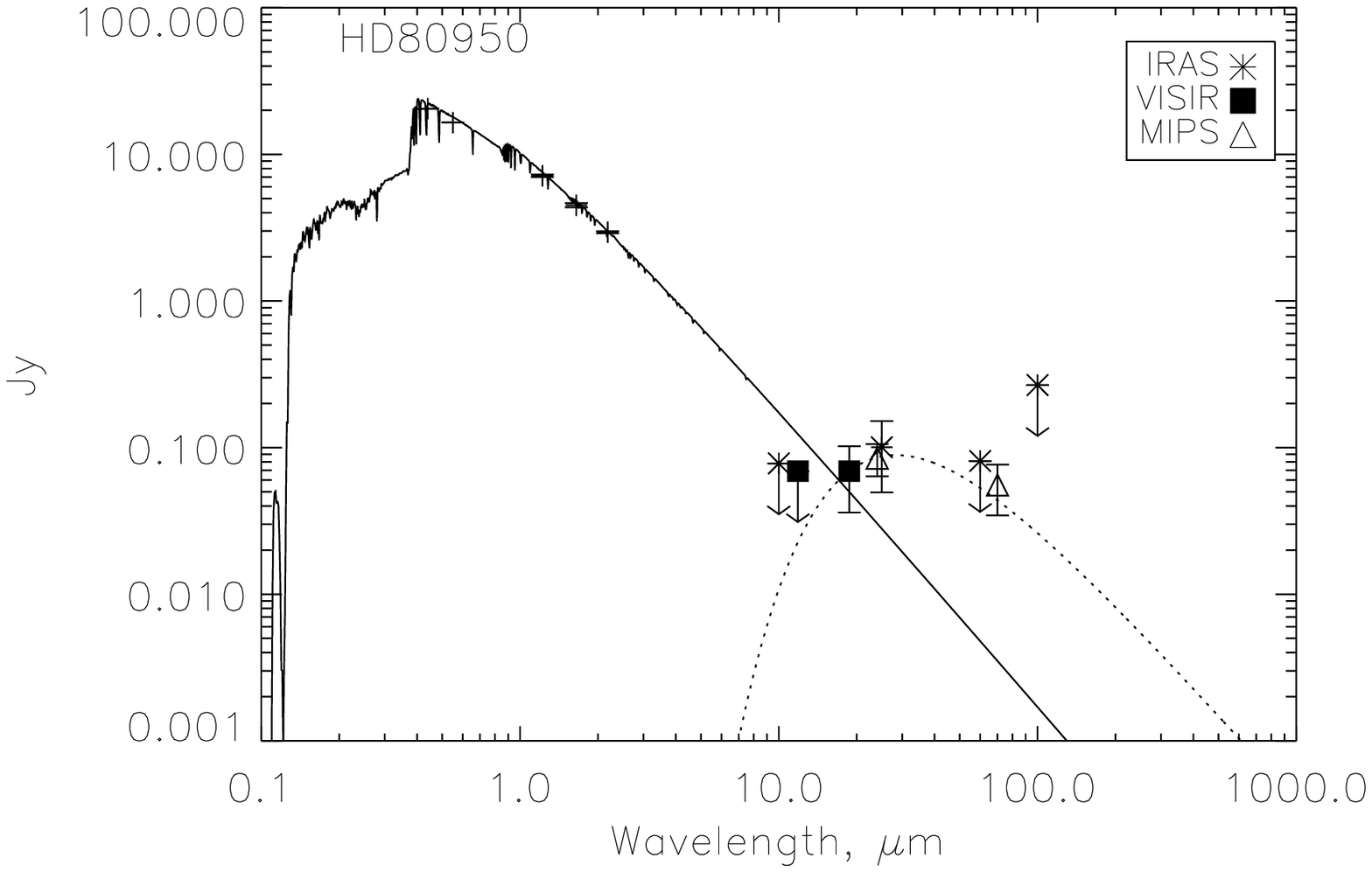}
\end{minipage}
\hspace{1cm}
\begin{minipage}{8cm}
\includegraphics[width=8cm]{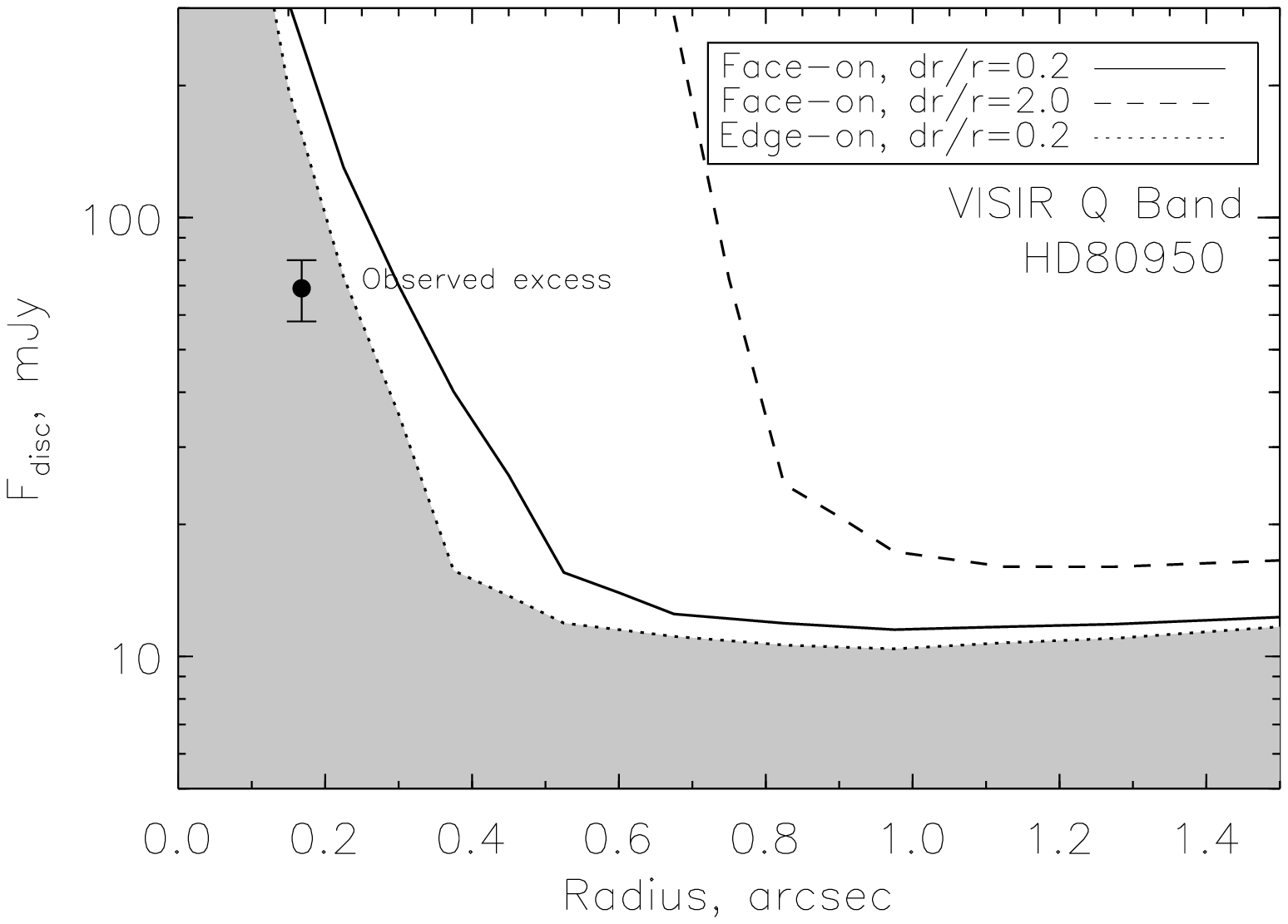}
\end{minipage}
\caption{\label{confplots} The observational results for sources with
  excess emission photometrically confirmed by our observations.  Top:
  The results for HD 3003.  The SED (left) includes a fit to the excess
  emission at a temperature of 265K (dotted line). The dashed lines
  indicate the photospheres of the two A star components, with the
  solid line being the sum of their emission.  Bottom: The results for HD
  80950. The excess emission is fit with a temperature of 180K
  (dotted line - solid line shows photospheric emission).  For
  both targets the limits on disc extension are shown in the right-hand
  plots, with the shaded region indicating the area the disc must lie
  in to have avoided being resolved.  The difference between
    the limits on disc extension arise from the differences in the PSF
    model for the different observations.  Specifically, the standard
    star observations associated with HD80950 showed large changes in
    the wings of the PSF over the course of the observations, giving
    rise to large uncertainties in the detection of excess emission
    extended over a broad spatial region.  Thus the limits on broad
    disc structures (dashed lines) with small central radius are
    poorer for HD80950 than for HD3003. The disc flux marked on these
  plots comes directly from the VISIR photometry.  The radial location
  marked arises from assuming the emitting material is blackbody-like.
  These results are discussed in detail in section 4.1.}
\end{figure*}

To test the limits we can place on disc extension with unresolved
images, we performed the same PSF subtraction and residuals testing on
models of stars + discs.  The stellar component was modelled as a
point source with flux set as predicted for each star in the
appropriate filter (see Table \ref{observations}).  Discs with
different radii, width and inclination to the line of sight and with
different levels of flux were added to the point source. The whole
model was convolved with the PSF as modelled by standard star images
to create a range of model images.  Different standard star images
were used to model the effects of PSF variation.  This process is
described in detail in \citet{smithhot}.  The disc geometries
  considered are simple ring-like discs with uniform brightness, with
  central radius $r$ and width $dr$.  The limits shown in Figures
  \ref{confplots}-\ref{plots754_141} are for disc widths $dr = 0.2r$
  (so a disc extending from $r-0.1r$ to $r+0.1r$ in radius) and
  $dr=2r$ (a disc from the central star to $2r$).  The central disc
  radius was varied from 0 to 1\farcs5 for the observations with VISIR
  and Gemini, and up to 4\arcsec for TIMMI2 observations.  The flux of
  the disc was scaled from 0 to 100\% of the total flux of the source
  in the observed filter.  Each model image was subjected to the same
  testing procedures as the science image itself, testing regions of
  the point-source subtracted image that had been optimised for the
  detection of extension for the disc parameters used as input.  These
  optimised regions were based on modelling work described in
  \citet{smithhot}. Regions above the lines for different disc
  geometries in Figures \ref{confplots}-\ref{plots754_141} 
represent disc models that were detected as extended objects (emission
detected in optimal testing regions) at a level of at least 3$\sigma$
(noise included pixel-pixel background noise and noise from PSF
uncertainty as detailed in \citealt{smithhot}). Regions below the
lines (shaded area) represent disc models that were not detected. 
The resulting limits on
detecting extended emission are dependent on PSF stability for discs
at small radii, and on the sensitivity of the observation for discs at
larger radii (see Figures 2 and 3 of \citealt{smithhot}).

As we do
not resolve any extended emission in the observations presented in
this paper, these limits are compared with the predicted radial
location of the disc.  The excess emission SED is fitted with a
single-temperature blackbody which is converted to a radial offset
assuming blackbody-like grains.  If the grains which dominate the
emission are small they will be inefficient radiators, hotter than
blackbody grains at a fixed radial location.  Thus the predicted
radial location will be an underestimate of the disc offset if small
grains dominate the emission (as has been seen in scattered light and
thermal imaging of resolved discs, see section 5).  
\citet{schneider} showed that for the HD181327 system its disc
  was imaged at a radius corresponding to 3 times that expected from a
  blackbody fit to the emission spectrum, a fact attributed to the
  emission from this disc being dominated by small inefficiently
  emitting but efficiently absorbing grains.  Other resolved discs
  have been shown to have radii that can differ from the blackbody fit
  by up to a factor of 3 (see Table \ref{tab:trueastars}).  As the
  grain properties of the discs are unknown, 
  in the results section the extension limits are compared to the
  radius suggested by assuming blackbody grains and up to 3$\times$
  the blackbody radius.  In all cases the extension limits are
  consistent with a disc lying at the blackbody radius (see Table
  \ref{results}) but the discs could also be dominated by smaller 
  grains at a larger radial offset (up to the extension limit). The
  exception in the single disc case is HD71155, for which the
  extension limits indicate the disc must have multiple belts
  (see discussion in section 4.2).  We
  therefore also determine the minimum grain size that will not exceed
  the temperature fit to the excess emission when at the extension
  limit listed in Table \ref{results}. This calculation requires the
  assumption of grain composition which we take for reference to be 
  non-porous grains  with no ice inclusions with a silicate
  fraction of 1/3 and 2/3 organic refractory material
  (see \citealt{wyattdent} for details of how grain
  temperatures were calculated for the assumed composition). These
  minimal grain sizes are listed as $D_{\rm{grain}}$ in Table
  \ref{results}. In all cases these grains are smaller than the
  blowout limit (1.3$\mu$m for an A0V-type star assuming the grain
  properties given above, or greater for cooler stars), and so it's
  more likely that the true disc radius is smaller than the extension
  limit given in Table \ref{results}.

%

\section{Results}
\label{s:res}

We split our debris disc targets into 3 subgroups based on the
observational results: those with excess independently
confirmed in our photometry; those for which limits can be placed on
the extent of the disc with these observations; and those for which we
cannot place limits on the disc with our data (HD 31295 and HD 38206).
These categories are identified in Table \ref{results}.  In addition
HD 23432 was found to have excess due to a reflection nebula and not a
disc.  This source together with HD 31295 and HD 38206 are described in
Appendix A.  The photometrically confirmed sources and those for which
we can place limits on the disc with our observations are described
below. 

\subsection{Photometrically confirmed discs}

\emph{$\eta$ Tel:} The excess emission towards $\eta$ Tel was
confirmed in our N band photometry with TIMMI2 at a 4$\sigma$ level of
significance (Table \ref{observations}).  Resolved imaging of this
target, revealing a two-component disc system with an outer component
lying at 24AU resolved at 18$\mu$m and a further unresolved inner
component which SED fitting shows is at 4AU, is presented in detail in
\citet{smitheta}.  We shall not discuss this source further in
this section, but will include this source in the discussion of the
sample in section 6.

\emph{HD 3003:} This star was identified as having significant
25$\mu$m excess (see Table \ref{sample}) by \citet{oudmaijer}. This
star is a binary: both components are A stars with similar
luminosities ($L_A =13.1L_\odot,
L_B=10.0L_\odot$, \citealt{dommanget}). The B
component was listed as being at an offset of 0\farcs1, 143$^\circ$
East of North in 1925.  The last confirmed observation of the separate
components was in 1964 with the B component at an offset of 0\farcs1,
171$^\circ$ \citep{mason}.

The star was observed with TIMMI2 at N and Q with follow-up on 8m
telescopes in both
bands (see Table \ref{observations}). None of the observations
resolved the separate stellar components. Excess was confirmed
photometrically at Q (detected 244$\pm$27mJy, expected
113$\pm$2mJy from photosphere giving an excess of
  131$\pm$27mJy - uncertainty on photosphere taken from fitting Kurucz
  model profiles to K band flux$\pm$1$\sigma$ error, and uncertainty on
  the detected emission and on the photosphere were added in
  quadrature to give error  on excess), although calibration
uncertainty prevents confirmation 
at shorter wavelengths.  This is in good agreement with
\citet{smithps} who presented MIPS 24$\mu$m confirmation of excess
for this source (detected 155 $\pm$ 9 mJy after subtraction of the
photosphere).  The recent 24$\mu$m and 18$\mu$m detections presented
here suggest a lower excess than that suggested by IRAS photometry (275
$\pm$ 20 mJy at 25 $\mu$m). Though the results are different at the
4$\sigma$ level only, it is possible that the emission from the
source may show temporal variance.  A dust temperature fit of
265$^{+30}_{-60}$K is consistent with the observed excess and shorter
wavelength limits (see Table \ref{results} and SED in Figure
\ref{confplots}).  Uncertainty on the blackbody temperature fit
  was taken from temperature range that allows a fit to within 3$\sigma$
  for all excess measurements.  Undetected background targets within
the TReCS field-of-view (section 3) are constrained to $<$10mJy at N,
confirming the excess is centred on the star.  

\begin{figure*}
\begin{minipage}{8cm}
\includegraphics[width=8cm]{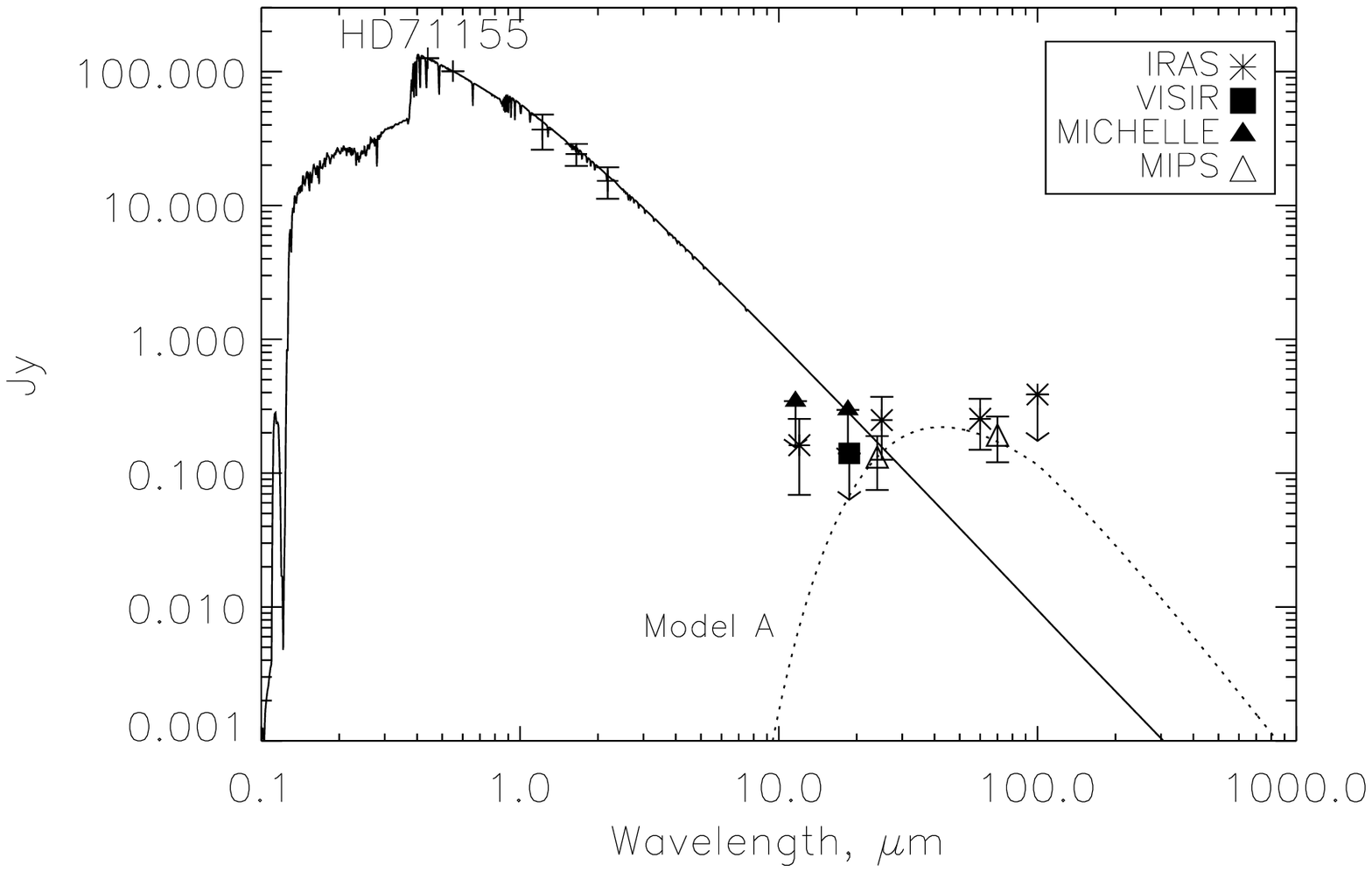}
\end{minipage}
\hspace{1cm}
\begin{minipage}{8cm}
\includegraphics[width=8cm]{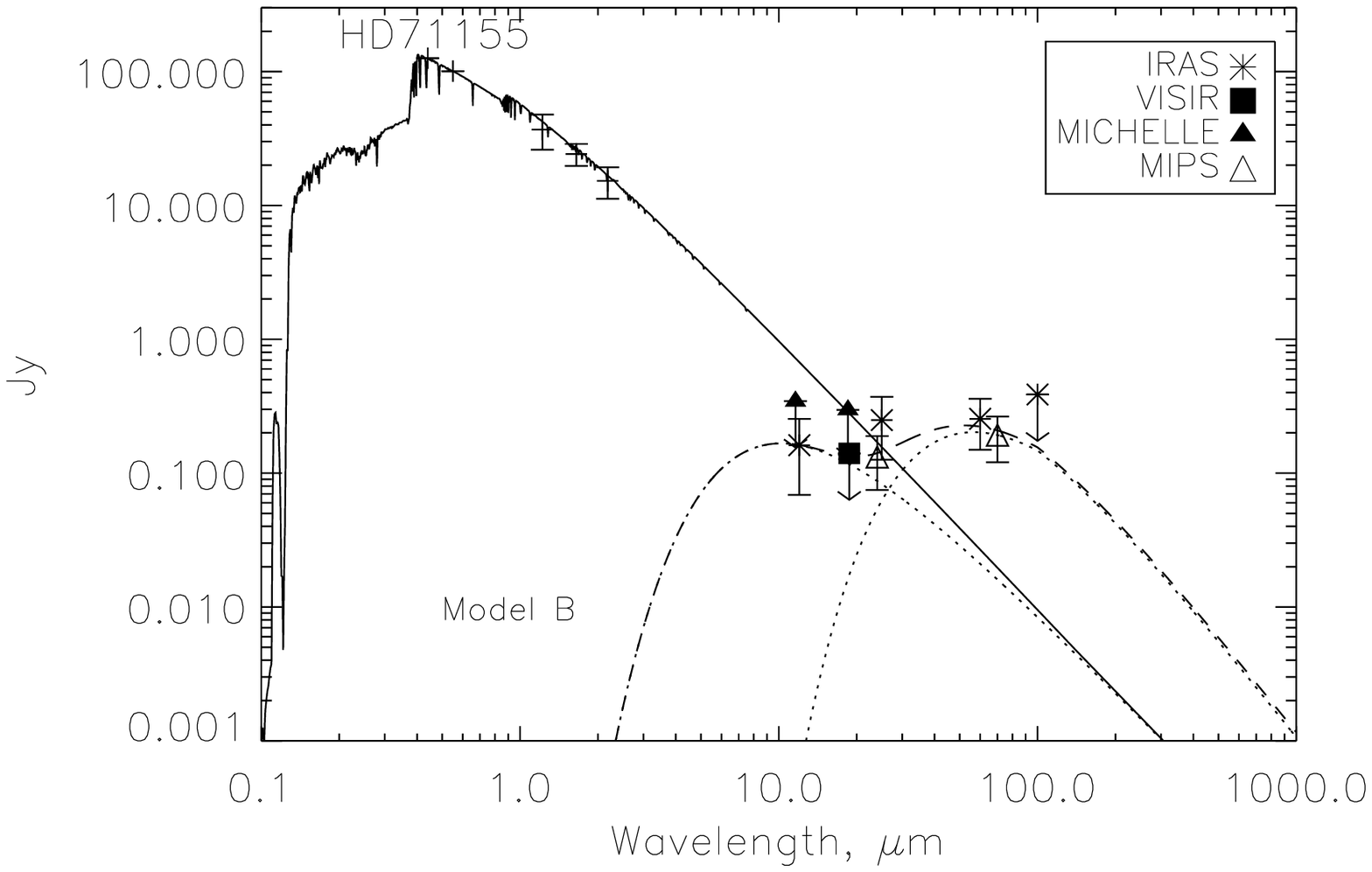}
\end{minipage}\\
\vspace{1cm}
\begin{minipage}{8cm}
\includegraphics[width=8cm]{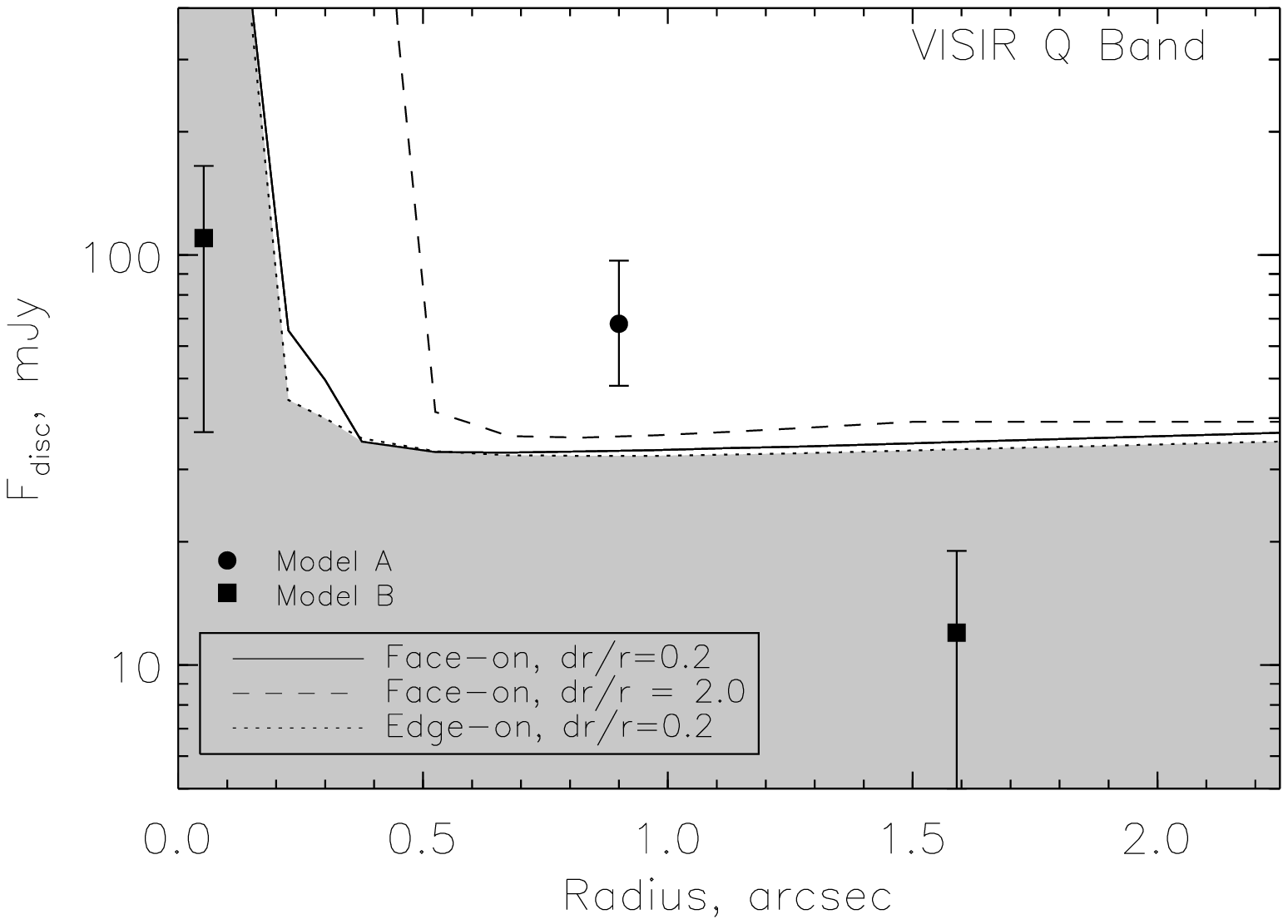}
\end{minipage}
\hspace{1cm}
\begin{minipage}{8cm}
\caption{\label{71155plots} Top left: The SED of HD 71155 with a single
  blackbody fit to the excess emission at a temperature of 120K
  (symbols $>5\mu$m are shown after subtraction of the photospheric
  contribution).  This suggests an offset of 34AU (0\farcs90) from the
  star.  This fit is marked on the extension limits plot left. The
  shaded regions are where the disc(s) is (are) expected to lie. The
  limits suggest an alternative two-disc model which also agrees well
  with the IRAS 12 $\mu$m data, as shown top right, is more
  likely. Values of $F_{\rm{disc}}$ are taken from the SED fits for
  each model, using scaling to the MIPS photometry and the
  uncertainties therein to give the errors.} 
\end{minipage}
\end{figure*}

The final images showed no evidence of extended emission at any band.
The Q band VISIR images place the tightest constraints on the disc
size, putting a limit of $<$6.5AU on the disc's extent.  If the binary
listed in \citet{mason} is a true binary at a separation of
0\farcs1 (4.7AU at 47pc, Table \ref{sample}) the stability analysis
carried out by \citet{holman} would suggest the disc cannot be
circumbinary, as the disc should be at a radius of $\gtrsim$ 0\farcs24
(11.3AU or greater if the orbital separation is larger or the orbit is
eccentric).\footnote{We have tacitly assumed here that the dust grains are
distributed following the orbits of the parent planetesimals.  As dust
grains are affected by Poynting-Robertson drag they may have a
different spatial distribution to the parent bodies. 
However, following the equations in \citet{wyattins05} we
find that the grain collisional lifetime ($t_{\rm{coll}}$) is shorter
than the Poynting-Robertson drag timescale ($t_{\rm{PR}}$) for all
disc radii up to the resolution limit, i.e. that
$t_{\rm{coll}}<0.04t_{\rm{PR}}$ for $r<6.5$AU.  The dust grains are therefore
likely to be collisionally dominated and occupy a spatial distribution
similar to the parent population.} We conclude that the disc must
therefore be circumstellar.  Assuming that the dust is around the
primary star the temperature fit of 265K translates to an offset of
4AU assuming blackbody grains using $r =
  (278.3/T)^2\sqrt{L_\star}$ (e.g. \citealt{backman}, where $r$ is the
  dust location in AU, $T$ is the temperture in Kelvin and $L_\star$
  is the stellar luminosity in units of $L_\odot$).  This assumes the
  grains are in thermal equilibrium with their environment.  Grains
  smaller than the wavelength at which the excess emission
    peaks are inefficient emitters and are thus hotter than blackbody
    temperature at that  radial offset from the star.  Such
  grains can be offset by three times the radius suggested by a
  blackbody approximation (see discussion in section 3.3).  
  Assuming a radius of 4 AU there are then two possibilities for
the system: either the dust is in a stable location and the binary
must have a semi-major axis of at least 14.4AU (or larger if the
binary orbit is eccentric, according to equations of
\citealt{holman}); or the binary is closer to the star and the dust is
unstable.  Such unstable dust populations have already been detected
in a small number of binary systems \citep{trilling}.  However, if the
dust is in an unstable region this could naturally explain the
tentative evidence for temporal evolution in the level of excess.  If
the dust resides in a stable location, the motion of the binary
changing the overall illumination of the system as it travels on its
orbit could also possibly explain any temporal evolution. Determining
the orbit of the binary will be a crucial step in determining the
stability of the dust in this system.

\emph{HD 80950:} This star was identified by \citet{mannings}
as a possible host of mid-infrared excess based on the IRAS 25$\mu$m 
measurement of its flux (excess 101$\pm$17 mJy, see Table
\ref{sample}).  The source was observed with TIMMI2 at N and VISIR at
N and Q, with the Q band photometry allowing a confirmation of the
excess (total detected flux 119$\pm$11 mJy, predicted stellar
photospheric emission in this filter 50$\pm$2mJy with
  uncertainty taken from 2MASS Kband uncertainty, see Table
\ref{observations}). A fit to the Q band excess emission and the 24
and 70$\mu$m excess reported in \citet{su} suggests a temperature of
180$^{+20}_{-30}$K for the excess emission (uncertainty
  determined by range of temperature fits that fit all excess
  measurements to within 3$\sigma$). \citet{morales} used a similar
blackbody temperature of 188K  to fit the Spitzer MIPS and IRS data on
this target.  No 
background/companion sources were detected in any of the images (see
Table \ref{observations} for brightness limits on such sources).  No
evidence of extension was detected in any of the images.  The
resulting limits on possible disc sizes and geometry are relatively
broad due to high levels of variation in the PSF during the
observations. For face-on discs the limits suggest a disc
radius of $<$24.5AU if the dust is distributed in a narrow ring (at a
distance of 81pc, Table \ref{sample}), or $<$61AU for a broad face-on
disc (see Figure \ref{confplots}). Assuming the grains are
blackbody-like suggests an offset of 13.6AU, consistent with these
limits for all disc geometries.  Grains much smaller than the
  peak of emission can lying at 3$\times$ blackbody radius (see
  discussion of HD 3003 and section 3.3) would lie at 40.8AU
(0\farcs51) which should 
be detectable in 8m observations with a more stable PSF.  With our
current constraints we can say that if very small grains dominate the
emission, the dust must be distributed over a broad radial range.   

\subsection{Limits on disc extension}

\begin{figure*}
\begin{minipage}{8cm}
\includegraphics[width=8cm]{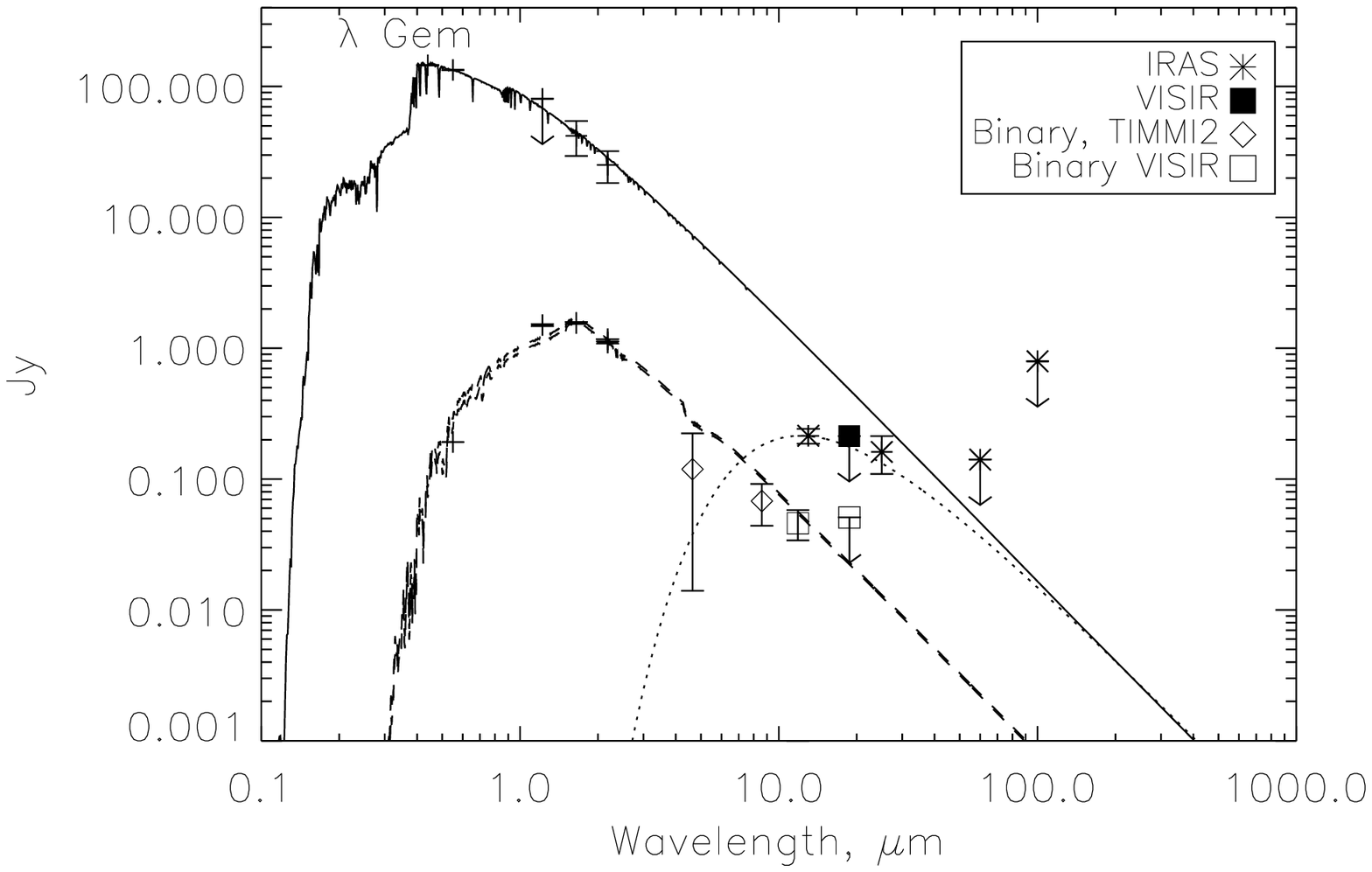}
\end{minipage}
\hspace{1cm}
\begin{minipage}{8cm}
\includegraphics[width=8cm]{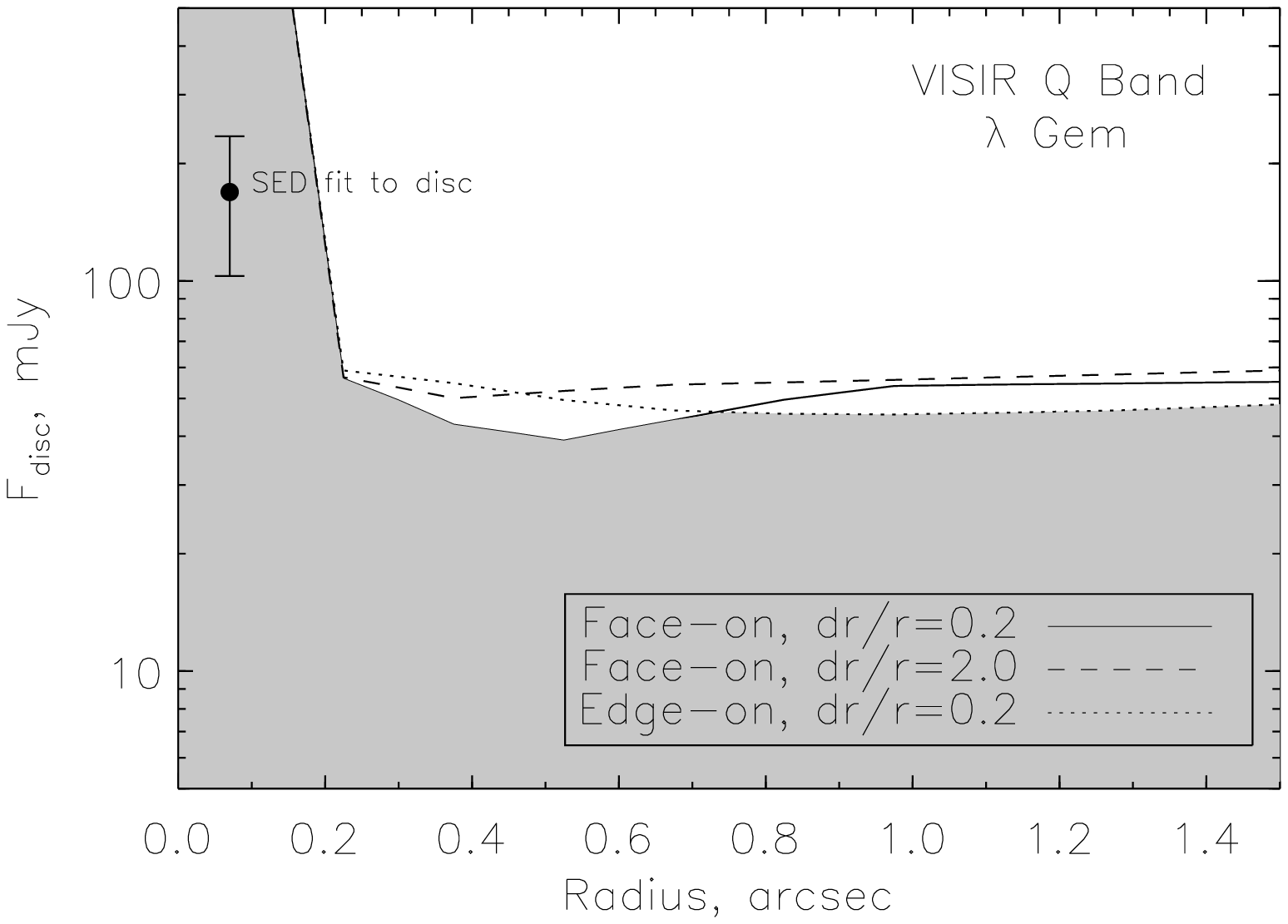}
\end{minipage}\\
\begin{minipage}{8cm}
\includegraphics[width=8cm]{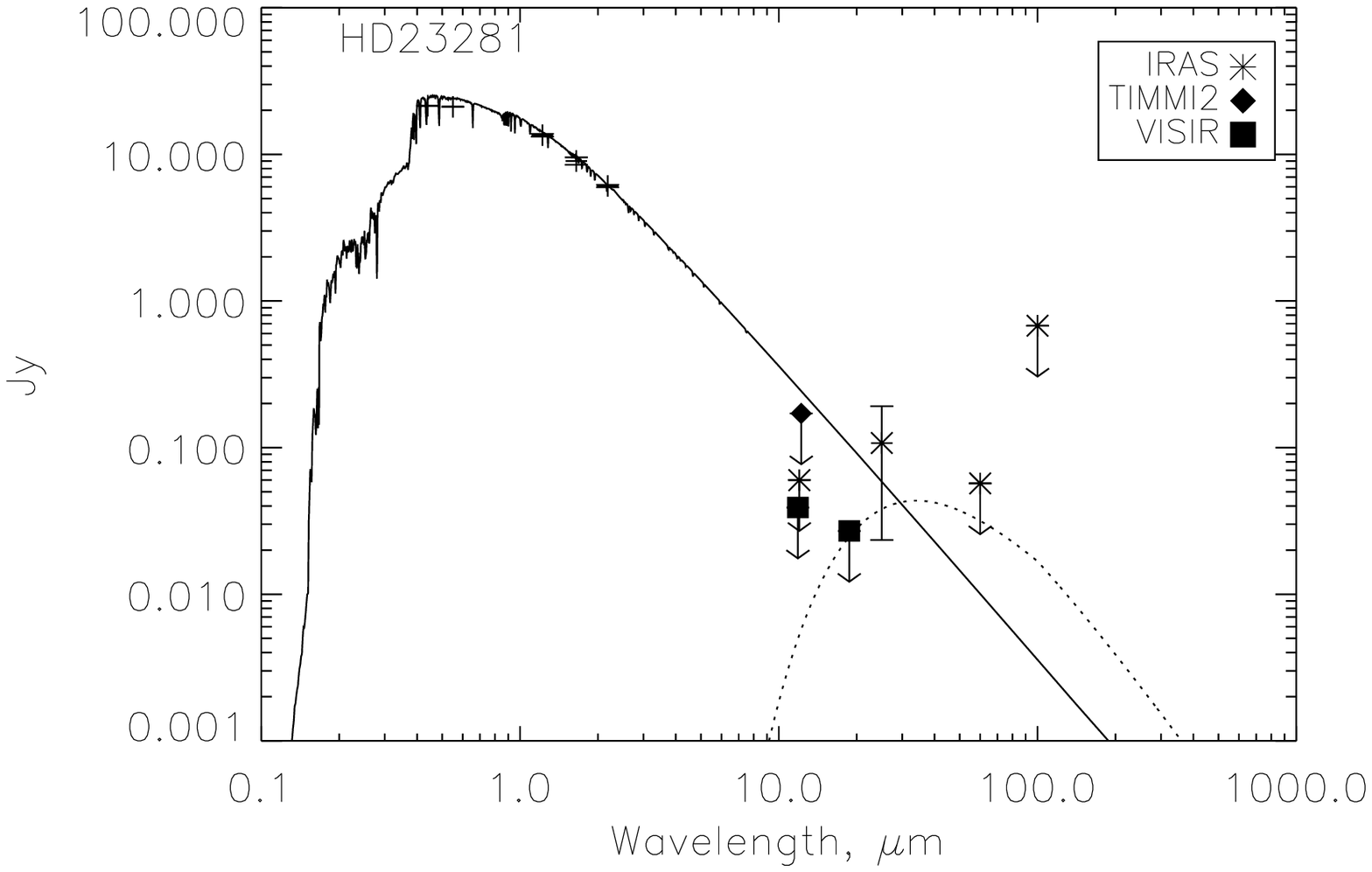}
\end{minipage}
\hspace{1cm}
\begin{minipage}{8cm}
\includegraphics[width=8cm]{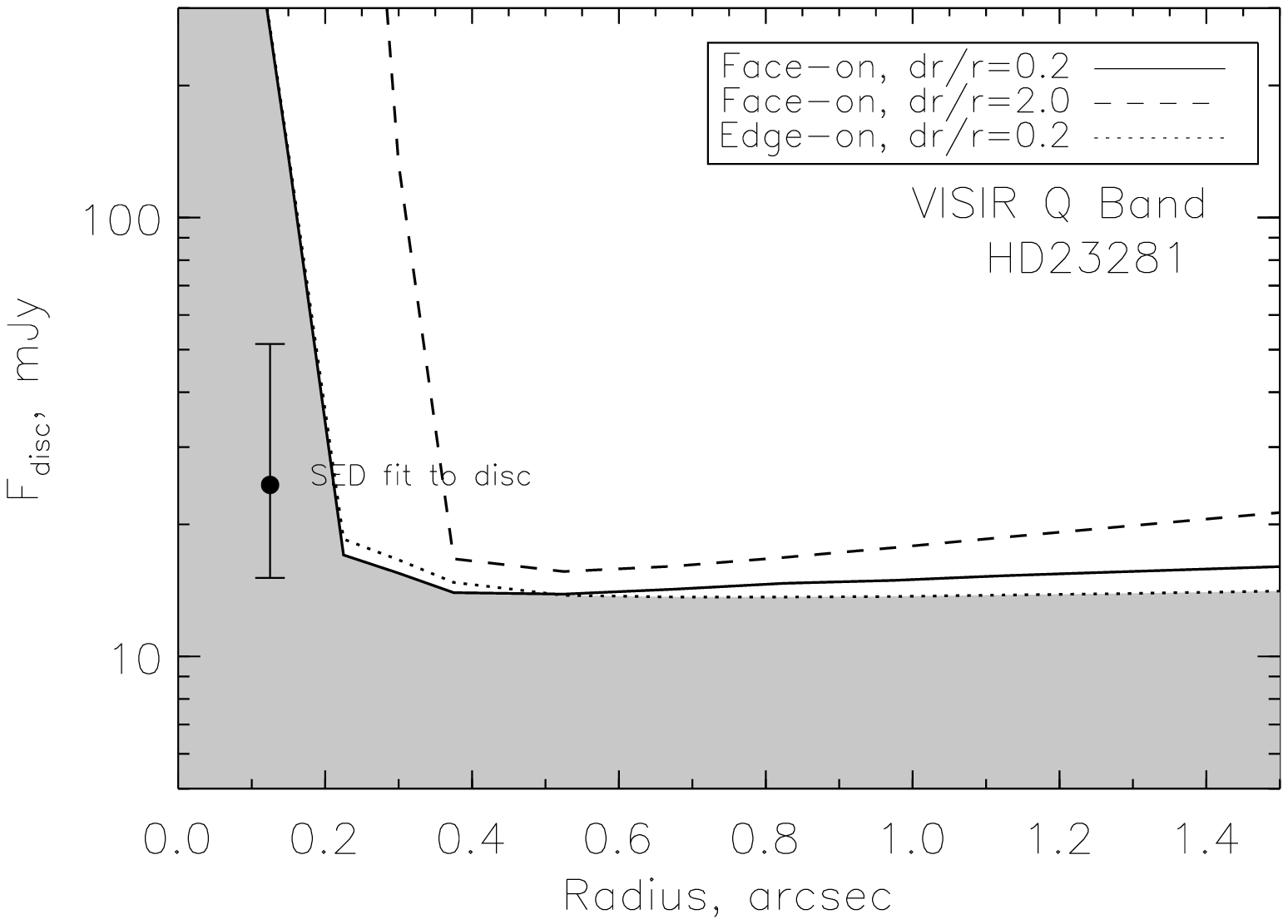}
\end{minipage}
\caption{\label{lam23281plots} The observations of $\lambda$ Gem and
  HD 23281. Top: The results of observations of $\lambda$ Gem.
  Bottom: The results of observations of HD 23281.  Symbols $>$5$\mu$m
  are shown after subtraction of the photosphere (not for the
  binary). Limits on extended emission obtained from the imaging are
  shown in the left-hand plots.  The shaded region marks the area that
  a disc would not be detected as extended emission.  The disc
  location marked on these plots is taken from the SED fit for disc
  flux, and from assuming that the dust is blackbody-like to determine
  a radius.  }
\end{figure*}

\emph{HD 71155:} This star was identified by \citet{cote} as
being a host of mid-infrared excess based on IRAS observations 
(see Table \ref{sample}). Calibration uncertainty
prevented photometric confirmation of the excess in our observations,
but we can rule out background objects within the TIMMI2 field of view
of $>$98mJy at N or within the VISIR field-of-view $>$16mJy at Q
(Table \ref{observations}). We thus confirm any excess should lie on
the target.  \citet{rieke} and \citet{su} used MIPS photometry to
confirm the excess at 24 and 70$\mu$m. These results suggested
somewhat lower excess than found with IRAS, but the difference is not
significant at 3$\sigma$.  In the following analysis we retain the
MIPS 24$\mu$m result (detected 302$\pm$19mJy) in preference to the
IRAS 25$\mu$m result to take advantage of the reduced errors.

The more recent measurements (upper limits at N and Q presented in
this paper and the MIPS excesses) allow a single temperature dust fit
(model A) at 120$\pm$15K (uncertainty from all temperatures that fits
excess data with 3$\sigma$, see Table \ref{results} and Figure
\ref{71155plots}; temperature similar to the 105K fit suggested by
\citealt{su}).  However, if the 12 $\mu$m excess detected by IRAS
is taken into account, which is not ruled out by the limits
presented here, then a two-temperature dust model (model B) fits the
spectral energy  distribution better (500$\pm$180K and
  90$\pm$20K, see Figure 
\ref{71155plots}).  Although no evidence for extension was seen on any
of our images, the non-detection in the VISIR Q band image allows us
to place constraints on the dust location in the context of the two
alternative models making the assumption that the different
temperatures represent different radial locations.  Figure
\ref{71155plots} shows that we should have detected resolved emission
if the dust was located at a radius of 34AU (assuming blackbody grains
at 120K; model A), regardless of disc geometry.  The results
therefore support a two component disc model, as was found for $\eta$
Tel \citep{smitheta}.  The inner component of this model is limited to
$<$8.4AU assuming a face-on orientation (Figure \ref{71155plots}). The
predicted location of 500K dust assuming blackbody-like grains is
$\sim$2AU.  This predicted location agrees with the results of
\citet{moerchenastar}, who found extended emission around this source
at 10.4$\mu$m consistent with a disc at 2AU. The outer component is
predicted to be at 61AU (1\farcs59) where sufficiently deep
observations on current instruments could resolve this
disc. 

\emph{$\lambda$ Gem:} This source was listed in \citet{cheng}, a study
of main-sequence A-stars, as having an IRAS excess.  This source is
notable as having one of the largest 24$\mu$m excesses amongst older
stars (\citealt{rieke}; stellar age is 560Myr).  The star is
listed in the Washington Double Star Catalogue as having a visual
binary companion at a distance of 9\farcs6 at a position angle of
33$^\circ$ East of North. Additionally,  component A has a binary
companion confirmed through lunar occultation measurements
\citep{dunham, richichi}.  These
measurements show evidence for a binary orbit that changes the
companion's relative position significantly over 20 years (offset
45mas, PA 300$^\circ$ in 1977, offset 14mas, PA 120$^\circ$ in 1999).

\begin{figure*}
\begin{minipage}{8cm}
\includegraphics[width=8cm]{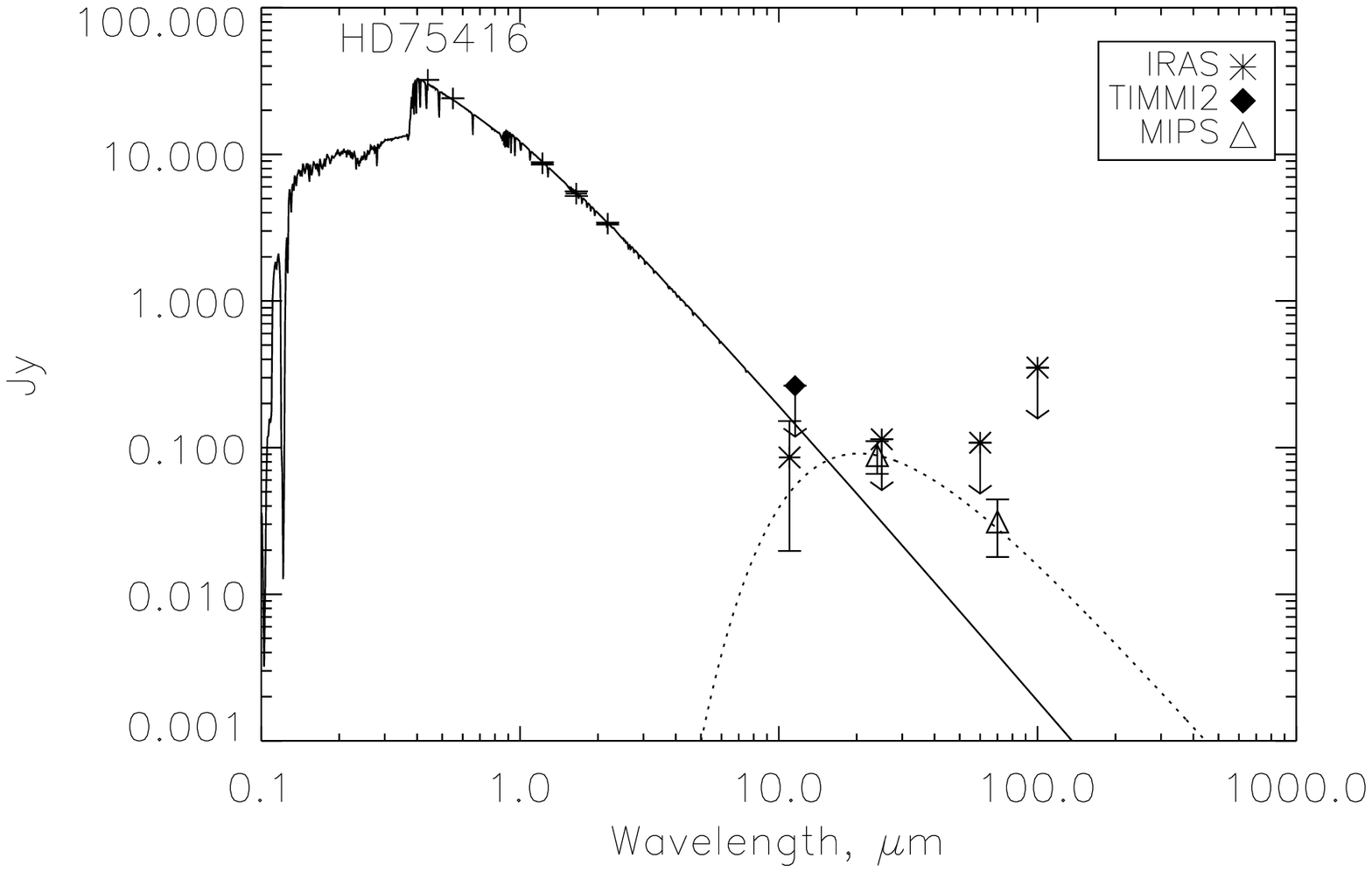}
\end{minipage}
\hspace{1cm}
\begin{minipage}{8cm}
\includegraphics[width=8cm]{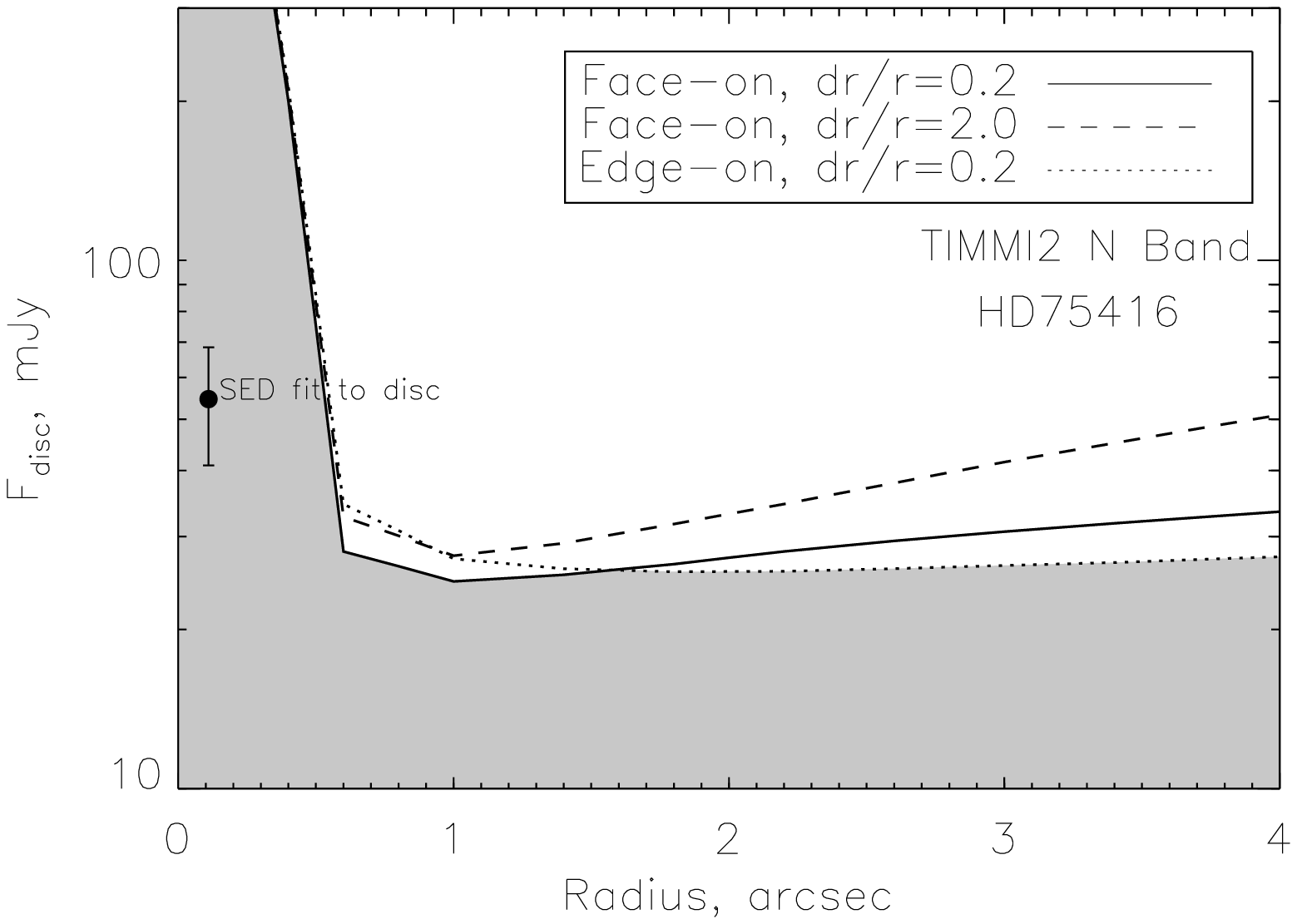}
\end{minipage} \\
\begin{minipage}{8cm}
\includegraphics[width=8cm]{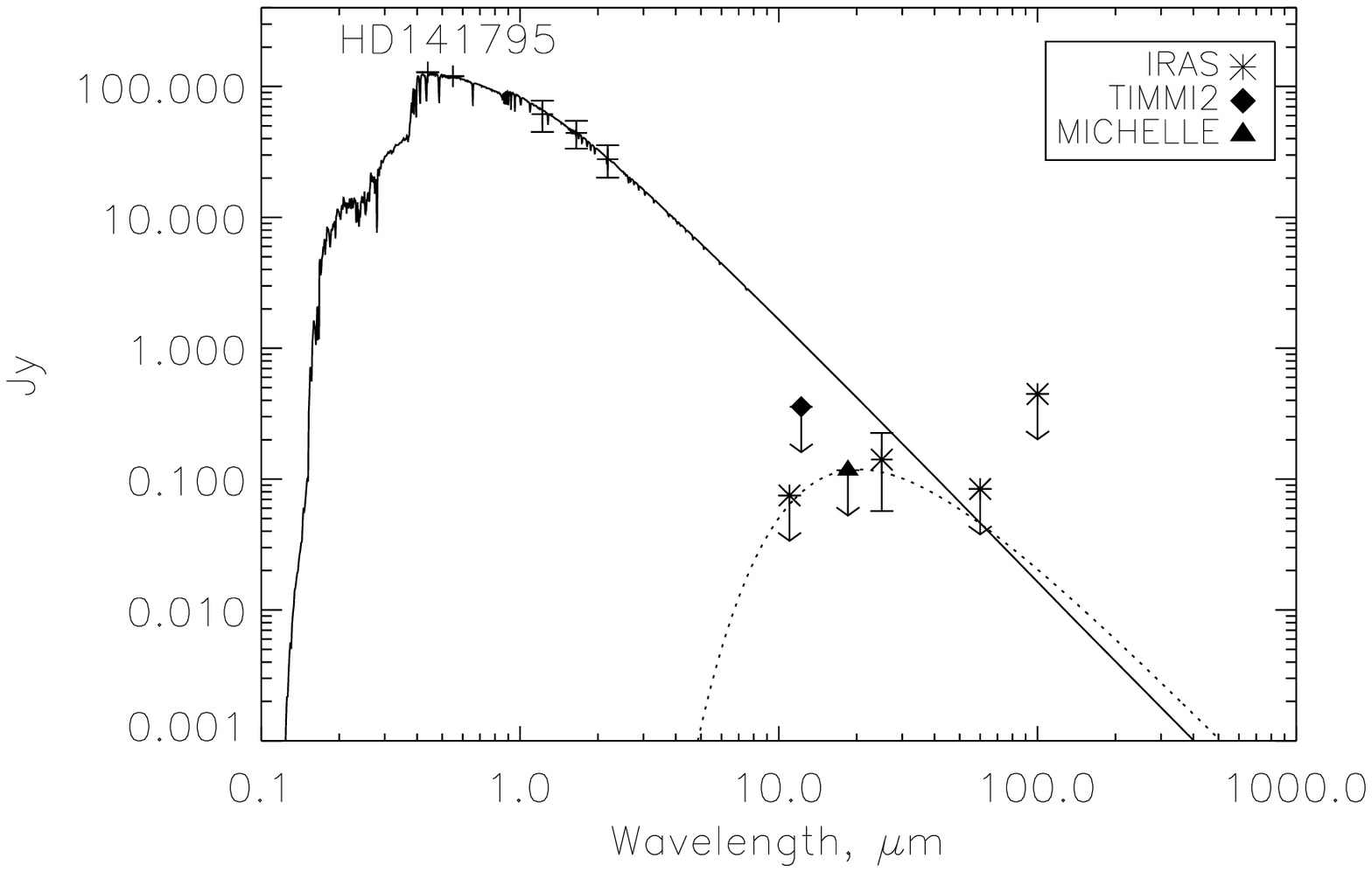}
\end{minipage}
\hspace{1cm}
\begin{minipage}{8cm}
\includegraphics[width=8cm]{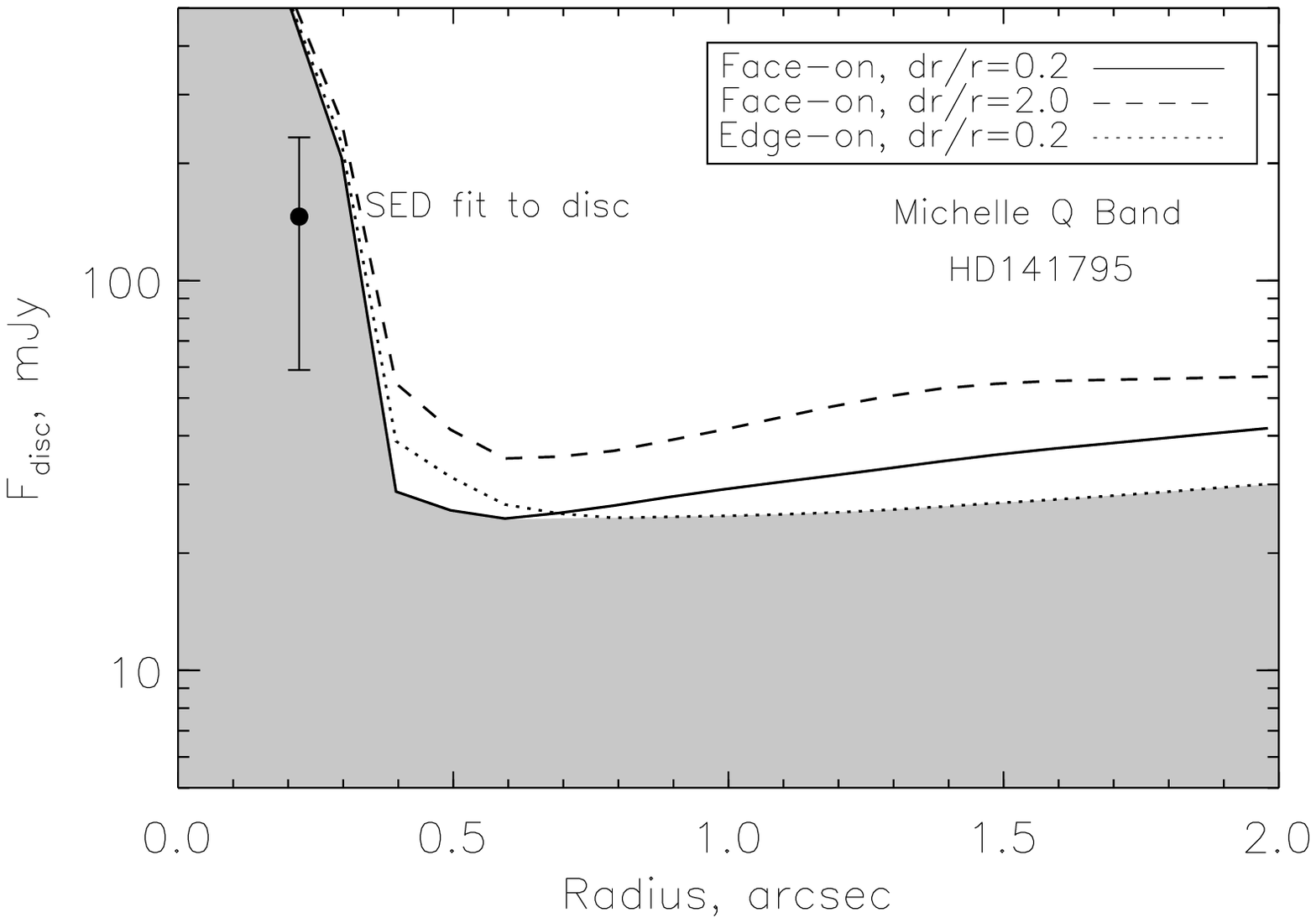}
\end{minipage}
\caption{\label{plots754_141} The results of analysis of the observations
of HD 75416 (top) and HD 141795 (bottom).  
Left: The SEDs with symbols at $>$5$\mu$m shown after
subtraction of the photospheric contribution at this wavelength. 
Right: The limits on disc location given by
non-detection of extension in the images. The grey areas show possible
disc locations for any of the geometries tried.  The filled circle
gives the predicted location from the SED fit. } 
\end{figure*}

The visual binary was resolved in TIMMI2 and VISIR N band observations at a
separation of 9\farcs83$\pm$0\farcs05, PA 30$^\circ\pm$1$^\circ$
(measured in VISIR image).  The secondary component is also
  resolved at M (central wavelength 4.6$\mu$m) with a flux of
  119$\pm$35mJy. We do not resolve the
separate components of visual component A. Calibration uncertainty
was high in the VISIR observations, as indicated by the low flux
measured on the primary (Table \ref{observations}), and thus we do not
photometrically confirm the excess. The flux of the
binary given in Table \ref{observations} and shown on the SED (Figure
\ref{lam23281plots}) was scaled to the expected primary photometry (so
multiplied by $F_{\textrm{primary, expected}}/F_{\textrm{primary, detected}}$).  The
visual binary was fitted with a K7 spectral type to fit the JHK
photometry from the 2MASS catalogue.  Adopting the parallax distance
to the primary of 29$\pm$2 pc the luminosity of the binary is
$0.12L_\odot$, consistent with a luminosity of $0.1L_\odot$ typical
for K8-type stars.

After subtracting the primary and binary contributions predicted from
the SED fitting the IRAS measurements still indicate significant
excess at 12 and 25$\mu$m (see Figure \ref{lam23281plots} and Table
\ref{sample}).  There are no nearby 2MASS or MSX sources likely to be
responsible for the excess measurements in the IRAS results, the
source is not in the galactic plane (b = 13.2), and no additional
sources are detected in our field of view, thus the IRAS excess
is not likely to be due to a background source (limits on background
sources listed in Table \ref{observations}). No evidence for
extension is found in the images.  We should have detected discs
larger than 6.1AU assuming a disc flux of $\sim$260mJy at Q from a
dust temperature fit of 420K (scaled to IRAS 12 and 24$\mu$m
photometry, error on the blackbody fit is 80K from errors on
  excesses measured in IRAS).  Blackbody
grains at this temperature would be at an 
offset of 2.2AU, or if the grains are small and lie at
  3$\times$ blackbody offset (as discussed in sections 3.3 and 4.1) they 
would be at 6.6AU, just beyond the limits on disc extension.  The
excess emission and SED fit should be confirmed before we can
interpret these limits in terms of constraints on the emitting
grains. 

\emph{HD 23281:} HD 23281 was first identified as a possible host of
mid-infrared excess by \citet{shylaja}.  IRAS photometry at 25$\mu$m
is indicative of excess emission at 3.8$\sigma$ significance (Table
\ref{sample}). Our photometric results do not confirm the excess on
this target (Q band photometry is consistent with excess but at a
$<3\sigma$ level, see Table \ref{observations}).  No additional
objects are seen in the field of view, with a limit on undetected
sources of $\leq 4$ mJy at N. There are no bright 2MASS or MSX sources
nearby that may have been caught in the IRAS beam and could be
responsible for source confusion.  There is no indication of extension
found on any of the images of this object.  Narrow discs at $\ge$9.5AU
or broader discs $\ge$15.9AU should have been detected as extended
emission in the VISIR Q band imaging (assuming $F_{\rm{disc}}=$ 24mJy
from SED fitting, see Figure \ref{lam23281plots}) .  We fit the IRAS
25$\mu$m photometry 
and VISIR upper limits on excess with a dust temperature of
210$^{+10}_{-80}$K . Blackbody-like grains at 210K would lie
at an offset of 5.4AU, 
consistent with the limits on extended emission. Very small grains
that could lie at 3$\times$ this offset (see sections 3.3 and 4.1) 
should have been detected  as extended
emission, if $F_{\rm{disc}}$= 25mJy although the level of excess and
dust temperature fit are currently too uncertain to allow constraints
to be placed on the dust properties. 

\emph{HD 75416:} HD 75416 ($\eta$ Cha) was identified as a possible
mid-infrared excess host by \citet{mannings} in their study of the
IRAS catalogues. It has significant excess at 12$\mu$m
(Table \ref{sample}). Our observation (TIMMI2, N band)
did not confirm the excess (detected 204$\pm$88mJy, predicted stellar
flux is 144$\pm$6mJy).  MIPS photometry at 24$\mu$m
\citep{rieke} is in good agreement with the IRAS detection at 25
$\mu$m (MIPS 128 $\pm$ 13 mJy photosphere 34$\pm$2 mJy, IRAS
117 $\pm$ 38 mJy photosphere 31$\pm$2mJy). \citet{su}
presented new 24 and 70$\mu$m photometry which also 
confirms the excess.  The limit on any background source within the
TIMMI2 field-of-view is $<$48mJy (3$\sigma$ at N), and thus it is very
likely the excess is centred on the source.  There is no evidence of
extended emission in the image, and thus we can place a limit on the
disc radius of $<$55AU on a disc of flux 55mJy in the N band
(based on a fit to the 
MIPS and IRAS photometry).  The fit to the excess at a temperature of
250K would translate to 11.1AU for blackbody grains, consistent with the
non-detection of extension in the TIMMI2 image (error on
  blackbody temperature 60K from errors on excess emission).  Even
very small grains at 3$\times$ the blackbody radial offset (see
section 4.1) would not be detected as extended emission.

\emph{HD 141795:} This star was listed as an excess candidate by 
\citet{shylaja}.  The IRAS 25$\mu$m measurements of this source's
photometry suggests an excess of 141$\pm$ 28mJy.  Calibration
uncertainties prevent a photometric confirmation of the excess in the
TIMMI2 and Michelle observations of the target (Table
\ref{observations}). Background and companion sources are ruled out
at a level of 49mJy (N band, TIMMI2) and 8mJy (Q band, Michelle), and
thus any excess is not likely due to detection of an additional source
in the IRAS beam.   

The images show no evidence for extension, and we place limits on
the extension of a disc with flux of 146mJy at Q (estimated from the 
25$\mu$m  excess measurement and 12 and 60$\mu$m upper limits) of
$<$6.2AU (Figure \ref{plots754_141}).  This is consistent with the SED
fit which uses a dust temperature of 250K putting the
dust at 4.2AU for blackbody grains. However, the true dust temperature
and disc flux at Q is highly uncertain (error on blackbody
  temperature fit is 70K).  Photometric confirmation of the excess is
necessary to confirm the limits provided by the extension testing. 


\section{Resolvability of discs in the mid-infrared}
\label{s:pred}

\begin{figure*}
\begin{center}{Disc resolvability : Gemini, 2 hours on source}
\end{center}
\begin{minipage}{9cm}
\includegraphics[width=9cm]{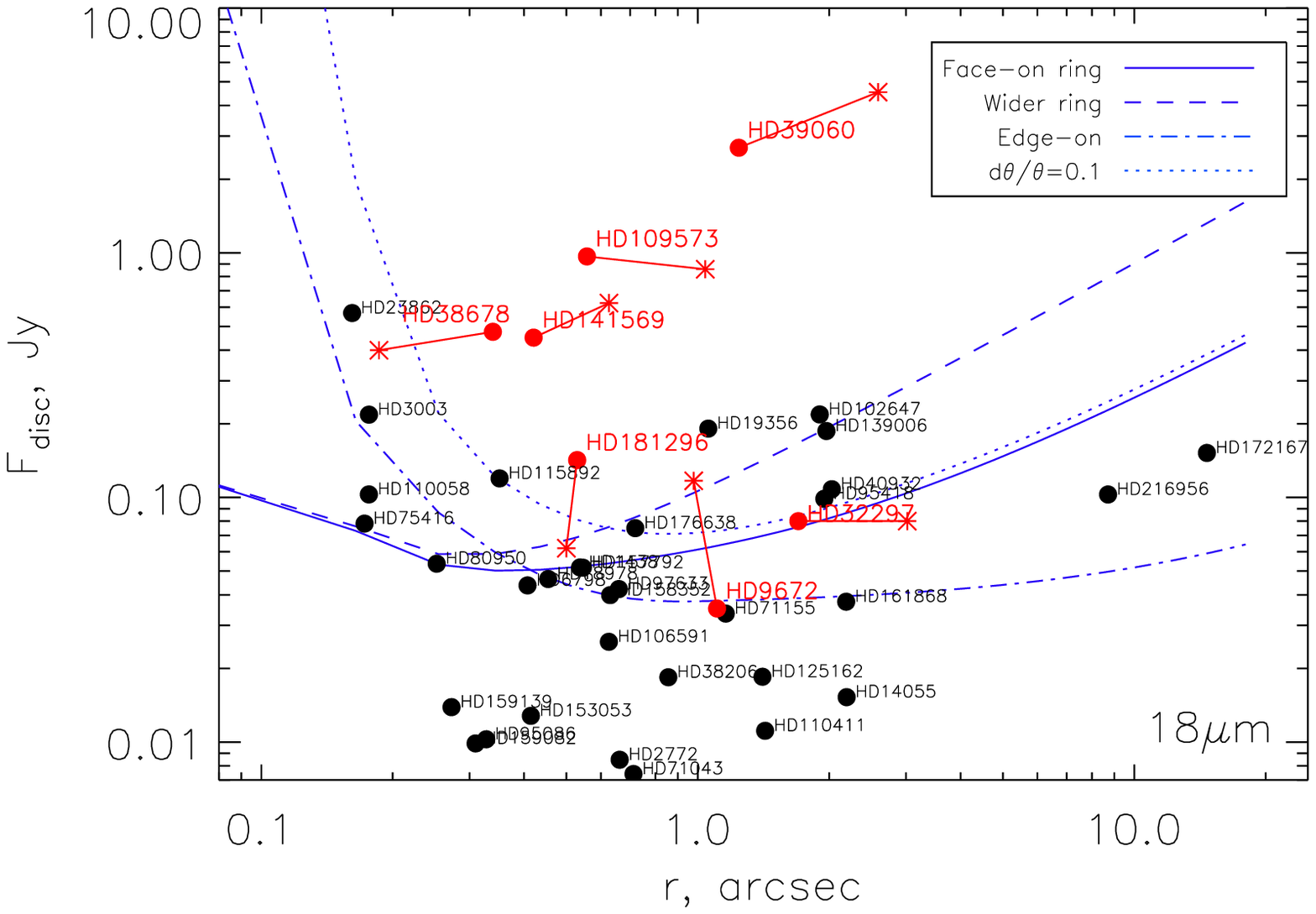}
\end{minipage}
\begin{minipage}{9cm}
\includegraphics[width=9cm]{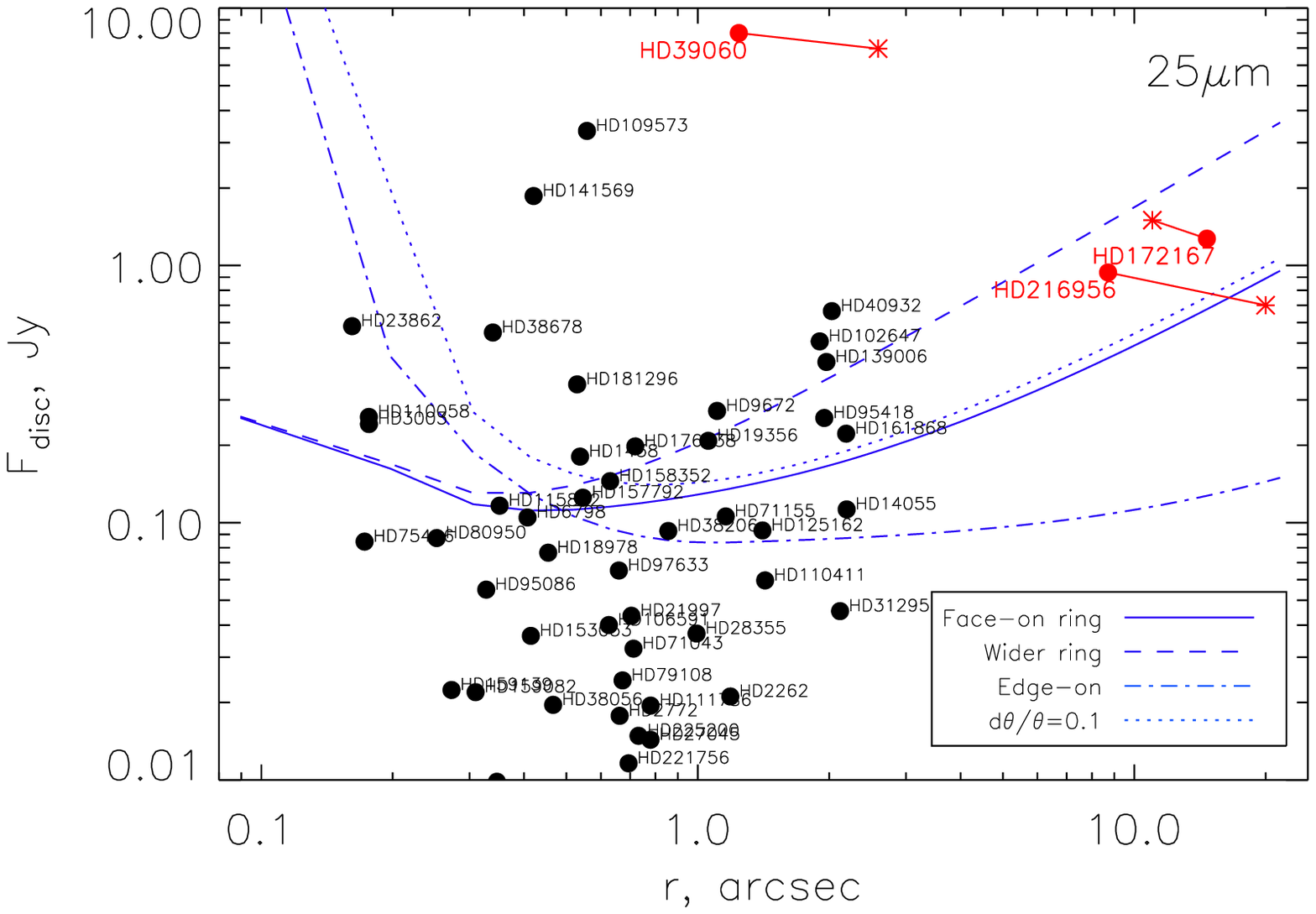}
\end{minipage} \\
\begin{center}{Disc resolvability : MIRI, 1 hour on source}
\end{center} 
\begin{minipage}{9cm}
\includegraphics[width=9cm]{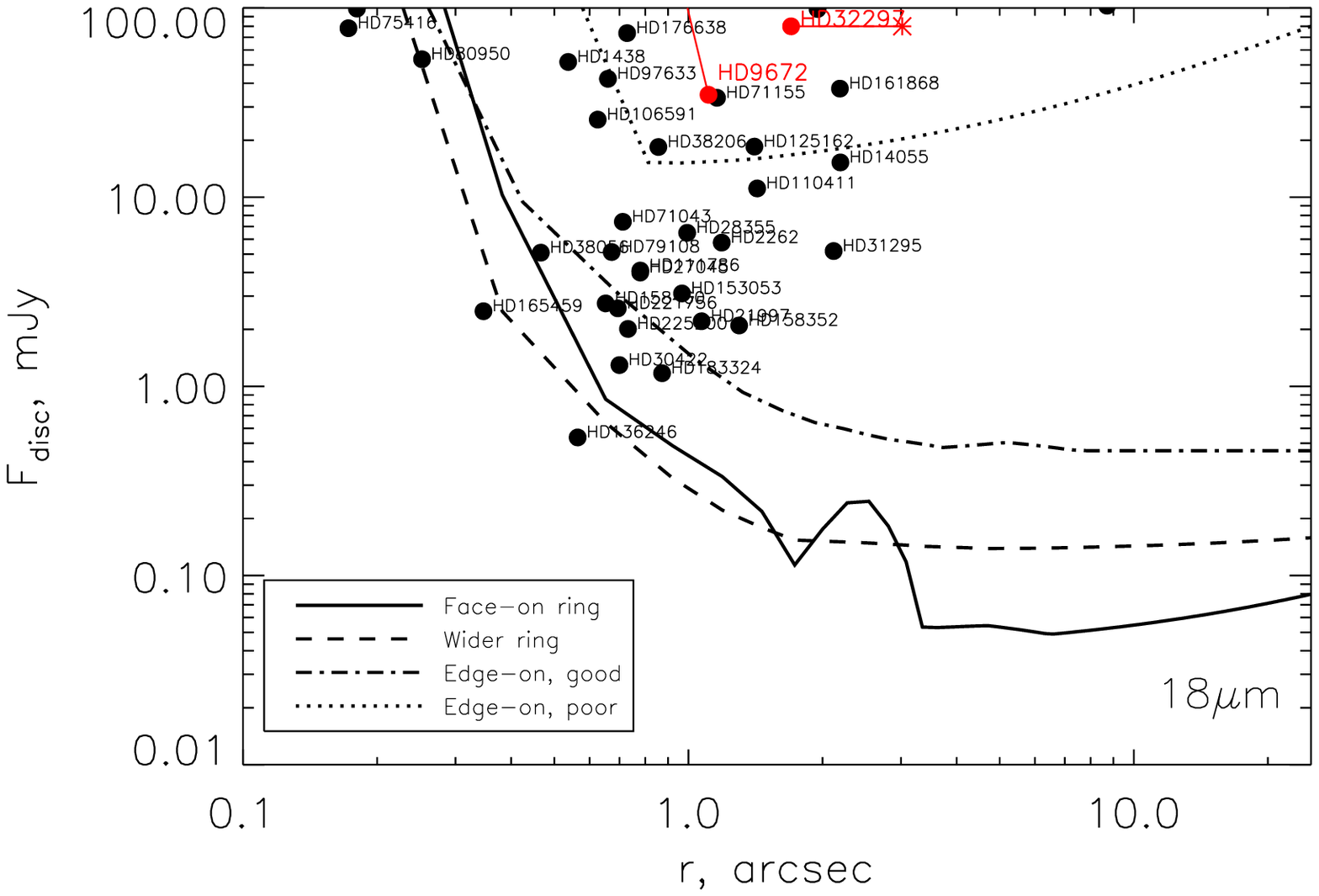}
\end{minipage}
\begin{minipage}{9cm}
\includegraphics[width=9cm]{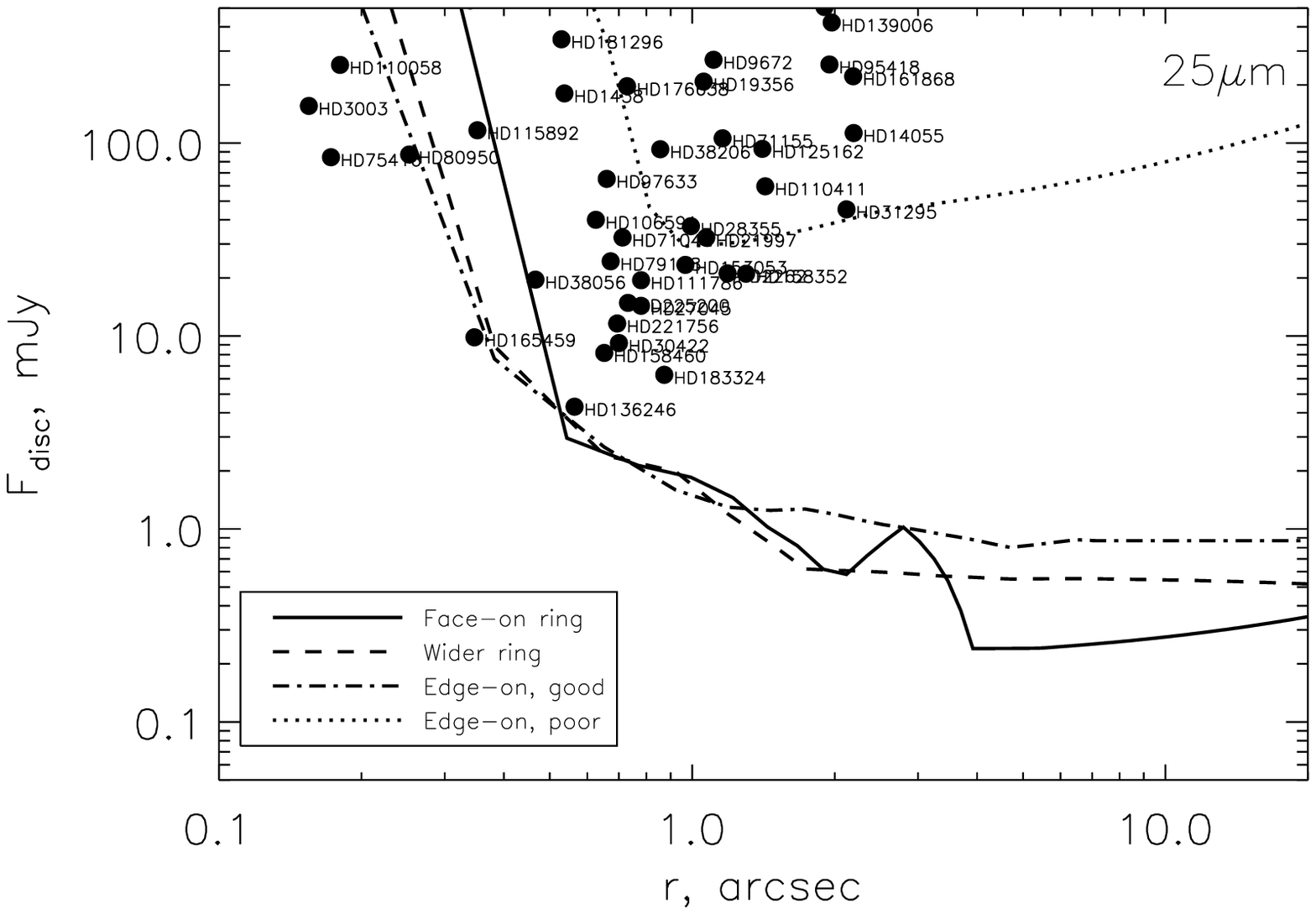}
\end{minipage} \\
\begin{minipage}{9cm}
\begin{center}{Disc resolvability : MIRI and METIS, 1 hour on source}
\end{center} 
\includegraphics[width=9cm]{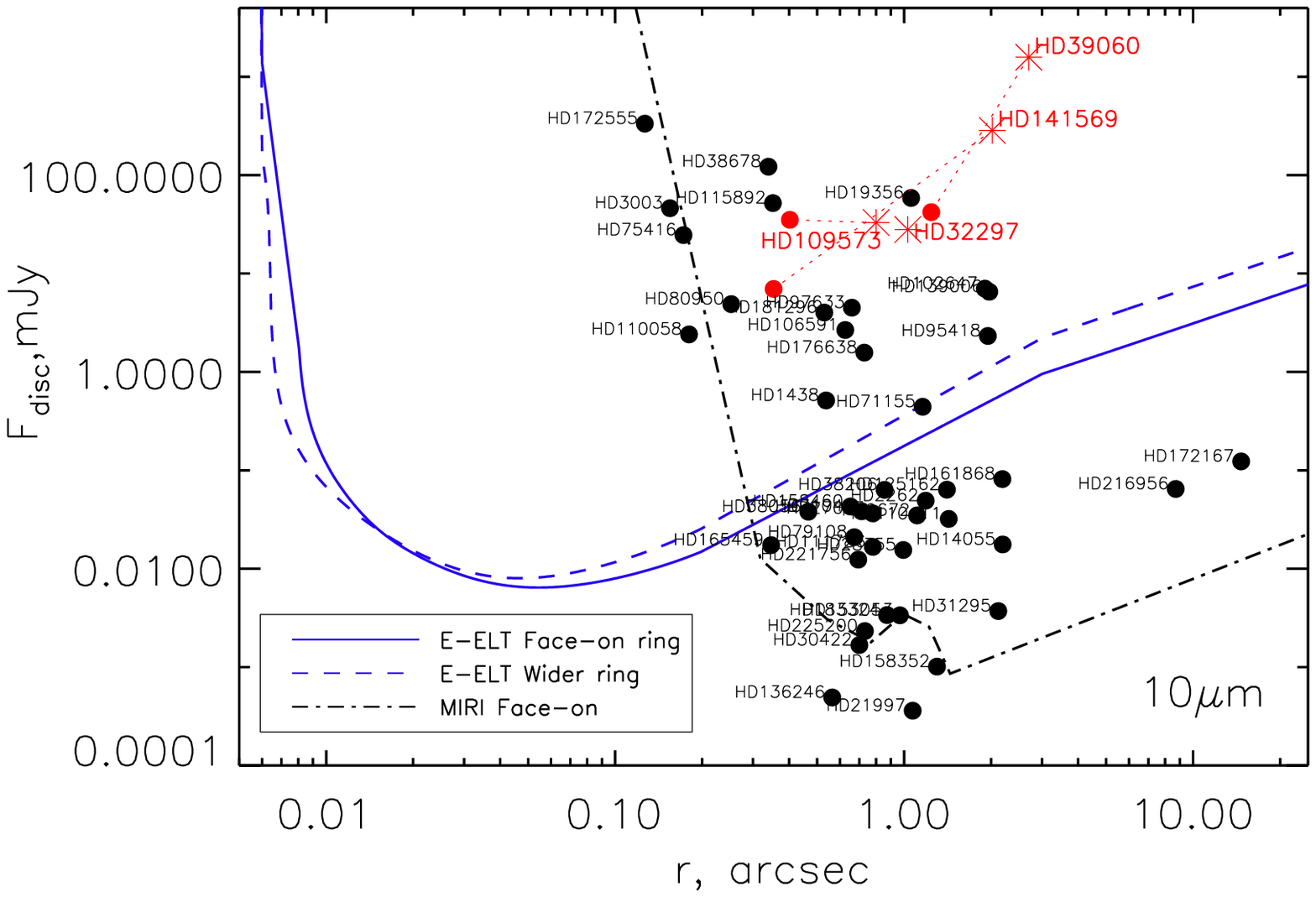}
\end{minipage}
\begin{minipage}{9cm}
\caption{\label{futurepred} Predictions for the resolvability of discs
with current and future instruments. See text for details of model
limits and disc properties.  Lines represent 3$\sigma$ detection
limits as described in section 3.3.  Different source geometries are
shown by the different lines and described in the legend (see also
section 3.3). Sources in red have already been resolved at this
  wavelength. Circles mark disc parameters estimated from SED
  fitting.  Asterisks mark the true disc fluxes and radial offsets
  seen at this wavelength for the resolved sources.  Errors
    from PSF variation for Gemini instruments and the E-ELT are
    approximated by using PSF model with FWHM varying from $\theta$ to
    $\theta + d\theta$ (see text for values of $\theta$).  This
    uncertainty is assumed to be at the 10\% level ($d\theta/\theta =
    0.1$) consistent with our 8m imaging data.  PSF variation is
    modelled by different simulated PSFs for MIRI.  We also show the
    limits that could be acheived for a perfectly stable PSF in the
    top panel (compare solid and dotted lines). }
\end{minipage} 
\end{figure*}

We now consider what mid-infrared debris discs could be resolved with
currently available instruments and future instrumentation. We use the
extension testing method described in section 3.3 and in detail in
\citet{smithhot} with PSF size and sensitivity appropriate to each
instrument considered to determine the limiting disc parameters for
resolution (disc size and flux for different geometries).  A short
description of the parameters used for each instrument considered are
given in the subsections below, and the limits shown in Figure
\ref{futurepred}. 

\begin{table*}
\begin{tabular}{*{8}{|c}|} \hline \multicolumn{2}{|c|}{Star} & Wavelength &
  \multicolumn{2}{|c|}{Predicted disc from SED fit}  &
  \multicolumn{2}{|c|}{Observed disc} &
  Reference \\ HD & Name & $\mu$m & Radius, \arcsec & Flux, mJy & 
  Radius, \arcsec & Flux, mJy &  \\ \hline 
  9672 & 49 Ceti & 18 & 1.42 & 35 & 0.98 & 68 & \citet{wahhaj} \\ 
  32297 & & 18 & 1.7 & 80 & 3.02 & 80 & \citet{moerchenb} \\
  38678 & $\zeta$ Lep & 18 & 0.43 & 475 & 0.18 & 400 &  \citet{moerchen} \\ 
  39060 & $\beta$ Pic & 18 & 1.59 & 2688 & 2.59 & 4336 & \citet{telesco} \\ 
  109573 & HR4796 & 18 & 0.51 & 1307 & 1.04 & 807 & \citet{telesco00} \\ 
  141569 & & 18 & 0.45 & 549 & 0.63 & 623 & \citet{fisher} \\
  181296 & $\eta$ Tel & 18 & 0.53 & 142 & 0.5 & 62 & \citet{smitheta} \\ \hline 
  39060 & $\beta$ Pic & 25 & 1.24 & 8005 & 2.59 & 6960 & \citet{telesco} \\ 
  172167 & Vega & 24 & 14.7 & 1271 & 11 & 1500 & \citet{su05} \\
  216956 & Fomalhaut & 24 & 8.7 & 939 & 20 & 700 & \citet{stapelfeldt} \\ \hline
\end{tabular}
\caption{\label{tab:trueastars} The predicted and measured disc
  parameters of sources with resolved debris discs in the
  mid-infrared. }
\end{table*}

On each plot we also show a representative sample of A star debris
discs for comparison with the determined limits (overplotted with
filled circles).  This sample is from \citet{wyattsmith07} and shows A
star discs detected at 24 and 70$\mu$m or 25 and 60$\mu$m.  Disc radii
and flux levels are taken from fits to the excess emission with a
single temperature blackbody as described in that paper.  There are
several uncertainties inherent in this fitting.  The radii
determined for the discs assumes the emitting grains behave like
blackbodies, when in reality small grains which are inefficient
emitters may dominate the emission and the dust could be up to 3 times
further from the star than this blackbody radius.
Multi-temperature fits to the excess are possible.
Different temperatures could represent different grain populations at
one radial location, or could indicate dust at
several radial locations (as is the case for $\eta$ Tel,
\citealt{smitheta}). In such cases the predicted radial location and
the level of disc flux for each component would be different from 
the simple single temperature fit shown here.  The level
of disc flux predicted at wavelengths other than 24 and 70$\mu$m (or
25 and 60$\mu$m) may be incorrect even in the case that a single
temperature is an accurate model for the emission.  This is
particularly true if spectral features are involved, for example
several spectra of debris disc targets with IRS on Spitzer have shown
strong silicate features in the N band \citep[see, e.g.][]{rieke,
  chen06, lisse09}.  The effect of these uncertainties can be seen in
the discs already resolved (shown in red in Figure \ref{futurepred}).
The resolved disc locations and fluxes are shown by asterisks and
listed in Table \ref{tab:trueastars}. The value of
$R_{\rm{obs}}/R_{\rm{pred}}$ (observed disc radius / predicted disc
radius) is as high as 2.3 (for Fomalhaut) for the restricted set of discs
resolved in the mid-infrared. This ratio is as low as 0.42 for
HD38678, which may have a multiple component disc (as suggested
by \citealt{fitzgerald}) which was incorrectly fitted with a single
  temperature.  As a final note
of caution, the population shown on these plots is only a sample of
known discs detected at 24 and 70$\mu$m and thus may not be truly
representative of the population of discs at 10 and 18$\mu$m (e.g. hot
discs detectable at 10$\mu$m may not be detected strongly at
70$\mu$m).  Discs around Sun-like stars, which will in general be
smaller than the A star discs (as dust must be closer to cooler stars
to heat to mid-infrared temperatures) are also excluded from the
sample shown.  These plots can be used as a
guide to the best sources to include in future observational
programmes aimed at resolving mid-infrared discs, but only through
such resolution can the true disc parameters be known.

\subsection{Gemini instruments}

We consider the detection limits achieved in 2 hours of observing
at 18 and 25$\mu$m (2 hours on-source; after overheads and repeated
standard star observation to monitor the PSF total observing time
approximately 8.5 hours).  The PSF model used is a Gaussian with FWHM of
0\farcs6 at 18$\mu$m (typical of 18$\mu$m observations presented here)
and 0\farcs72 at 25$\mu$m (taken from \citealt{telesco}). The point
source sensitivity follows from the detection levels found in the
0\farcs5 apertures at 18$\mu$m and is 1.8mJy in 2 hours on source;
extrapolation to 25$\mu$m (sensitivity 4.8mJy in 2 hours) assumes a
factor of 8/3 brightness increase needed for a source to achieve the
same signal-to-noise in the Qb filter (25$\mu$m) of TReCS as in the Qa
filter (18$\mu$m), as outlined on the Gemini website.  The
  detection limits for extended disc emission were determined in the
  same way as the extension limits for the observations presented in
  this paper. Models of disc+star emission (with disc geometries and
  disc flux as described in section 3.3) convolved with the model PSF
  were treated as model images, and subjected to extension testing.
  Point-source subtracted model images were tested for significant
  residual emission in optimal testing regions as described in
  \citet{smithhot}.  Emission above 3$\sigma$ significance was
  regarded as a detection of extended disc structure. 
The resulting
limits (Figure \ref{futurepred}, top line) show that the best targets
for resolved disc imaging campaigns are those that have already been
resolved (as for Figures 1-4, the region above the lines
  represents the disc parameter space that would result in a
  significant detection of extension according to the method outlined
  in section 3.3 and in detail in \citealt{smithhot}).  This plot was
used to identify $\eta$ Tel, whose excess was 
independently confirmed in the TIMMI2 data presented here, as a prime
target for 8m resolution.  The resulting observations presented in
\citet{smitheta} resolved the outer disc component and highlight the
utility of this technique.  Of the known A star debris discs
population few sources remain that could be resolved in reasonable
observing times with current instruments at 18$\mu$m (those most
amenable to resolved imaging are HD19356, HD139006 and HD102647 from
current predictions of the disc parameters).  More discs could be
resolved at 25$\mu$m, although conditions suitable for 25$\mu$m
observing are more rare. 

\subsection{MIRI on the JWST}

\begin{figure}
\begin{minipage}{6cm}
\includegraphics[width=6cm]{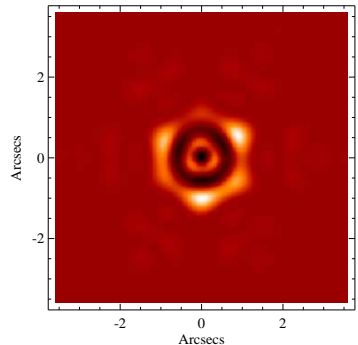}
\end{minipage}
\begin{minipage}{2.5cm}
\caption{\label{fig:miripsf} A model of the MIRI PSF at 18$\mu$m made
  using the JWPSF tool.  The model PSF is shown 
  after subtraction of the best fitting Gaussian to show the low level
  structural features resulting from the segmentation of the primary
  mirror. }
\end{minipage}
\end{figure}

The James Webb Space Telescope is due to be launched in 2013.  MIRI,
the mid-infrared instrument, will provide vastly greater sensitivity
to debris discs in the 5--27$\mu$m spectral range.  Since the  primary
mirror of the JWST consists of 18 hexagonal segments with the combined
collecting area of 25m$^2$ (similar to a 6m circular primary), the
shape of the mirror segments is reflected in the shape of the expected
PSF.  Thus the PSF was taken from the JWPSF tool for an A5 type star
(tool at http://www.stsci.edu/jwst/software/, see Figure
\ref{fig:miripsf} for PSF). Noise from possible PSF uncertainty was
calculated as the differences in residual flux emission found when
using different PSF models as determined using the different OPD files
available (files represent modelled mirror misalignment and
aberrations). We also tested PSF models based on A0 type stars and
found no significant difference in the limits achievable.  The optimal
regions for testing for extension as described in \citet{smithhot} and
used to determine the extension limits on sources in this paper (see
section 3.3) were modified to exclude pixels close where differences in the
PSF model from different OPD files were high ($\ge$50\% the level of the
signal of the PSF model in a pixel).  The point source sensitivities
were assumed to be 0.7$\mu$Jy at 10$\mu$m, 4.3$\mu$Jy at 18$\mu$m and
28$\mu$Jy at 25$\mu$m (10$\sigma$ in 10,000s) following the guidelines
on the JWST webpages (http://www.stsci.edu/jwst/science/sensitivity/).  
These sensitivities were converted to 1 hour's on-source
  integration by assuming that signal-to-noise varies as
  $time^{0.5}$.  Overheads have not been included in this
  calculation.

The resulting limits on resolved disc parameter space are shown in
Figure \ref{futurepred}, middle panel and bottom left for face-on
discs at 10$\mu$m.  The increase in sensitivity over current
ground-based imaging is clear from the low levels of disc flux for
which detection of extended emission is possible in only 1 hours
observing. Almost 100\% of the A star discs detected at 24 and
70$\mu$m should be resolvable with MIRI, although the resolution of
the discs close to the inner radius limit will strongly depend on
accurate PSF calibration (and of course the caveats relating to disc
flux / radius predictions from SED fitting must be considered). 
The resolution of edge-on discs is a strong function of
position angle as the PSF is not circularly symmetric.  Discs which lie
along the direction of the corners of the hexagonal shape (see Figure
\ref{fig:miripsf}) are more difficult to detect in residual emission
as these regions see more noise resulting from mirror misalignment and
aberrations; observing at different position angles could mitigate
against this issue.  The 'bump' in the detectability limits for 
face-on ring models also arises from the hexagonal feature of the
PSF.  Wider rings and discs lying edge-on are less strongly
effected by this as their detection depends on less confined radial
locations.  A four quadrant phase-mask (4QPM) coronagraph will be
offered at 3 wavelengths; 10.65, 11.4 and 15.5$\mu$m (wavelengths
optimised for planet detection, \citealt{boccaletti_miri}).  We approximate
the effect of including a 4QPM at 10$\mu$m (bottom panel, Figure
\ref{futurepred}) by increasing the sensitivity of the observations by
a factor of 250 at 0.3$\lambda$/D and 50 at 5$\lambda$/D, falling to a
factor of 1 at 10$\lambda$/D (where D is 6.5m).  These values are
based on Figure 8 of \citet{boccaletti_miri}.  Including these
  sensitivity improvements allows the detection of discs of flux
  down to a limit of $\sim$9$\mu$Jy if the disc is at the optimal
  detection radius (1\farcs44, minimum of dot-dashed line in Figure
  \ref{futurepred} bottom panel).  Without the coronograph the minimum
disc flux required for the detection of extended emission is
0.013mJy.  A Lyot mask optimised at
23$\mu$m  will also be provided, and will be used primarily for the
detection of cold circumstellar discs.  However, due to the large
opaque mask of the Lyot objects at $\le$ 1\arcsec cannot be detected
with the Lyot \citep{boccaletti_miri}. As we can explore within this radius
with our PSF subtraction method we do not include the effects of a
coronagraph in our predictions at longer wavelengths (middle panel,
Figure \ref{futurepred}). 

\subsection{METIS on the E-ELT}

The European Extremely Large Telescope (E-ELT) is currently planned to
have a 42m dish and to start operation in 2018. In the mid-infrared
the current proposed first generation instrument is METIS, which will
cover the 3-13$\mu$m range (although formerly Q band was proposed for
this instrument, \citealt{brandl}).  We scale our PSF models from 8m
instrumentation to a 42m dish giving a FWHM of 0\farcs06 at 10$\mu$m.
The sensitivity for imaging was assumed to be 8$\mu$Jy in 1 hour on
source (10$\sigma$) as taken from the METIS webpages
(http://www.strw.leidenuniv.nl/metis/). To approximate the effect of
the inclusion of a coronagraph we assumed that the sensitivity would
be improved by a factor of 10 at our hard limit of disc detectability
(0\farcs006 where we adopted a PSF variation of $0.1\times$FWHM which
is typical of the level of variation we see in our 8m imaging data).
This factor was taken from the increase in sensitivity predicted for
planet detection when including a four quadrant phase mask on top of
PSF subtraction as given in \citet{brandl}.  
For larger radii we assumed the sensitivity improvement would decrease
linearly with radius to a fixed level of 1 at $\ge$0\farcs5 ($\sim$8
times the FWHM of the assumed PSF). This
model is highly over-simplified and does not take into account many
issues such as how sensitivity improvements may differ from point
source predictions for extended emission . We note that work
currently being undertaken by Eric Pantin and collaborators to
determine the sensitivity of METIS to disc and planet detection will
adopt much more realistic PSF and coronagraph models.

The predictions for resolvable disc parameter space with METIS at
10$\mu$m are shown in Figure \ref{futurepred} bottom left.  The
comparison with MIRI on the JWST is representative of the different
strengths of ground-based and space-bourne instrumentation.  Larger
discs with lower surface brightness will be ideal targets for MIRI
observations, but a large dish like the E-ELT will be needed to
resolve very small discs ($\lesssim$0\farcs3), or indeed 
structure within larger discs.  Although the A star disc sample shown
on this plot does not fill much of the small scale region, this is
because discs detected at 24 and 70$\mu$m are cool and thus at large
radii.  The inner components of multiple dust population discs or dust
at 10$\mu$m around Sun-like stars will live in these small spatial
regions (see earlier discussion). This 10$\mu$m disc population is
poorly known as studies of these discs have been limited by
calibration accuracy (specifically we cannot detect discs by
 classical aperture photometry fainter than
the level of accuracy with which we know the stellar emission at this
wavelength, which is typically 10\% of $F_{\star}$, although
interferometric techniques can allow this limit to be surpassed).

METIS on the E-ELT would enable the discovery of disc populations
which cannot be detected photometrically (see above) through the
detection of extended emission at 10$\mu$m.  At the optimal
  detection radius (0\farcs05) METIS should be able to detect extended
  discs at the level of 6$\mu$Jy.   We can compare this value to the
  flux expected for an exozodiacal cloud around nearby stars. If a
  star has 1 zodi of emission between 0-3AU with constant optical
  depth then it is most likely to be resolved in observations with
  METIS at 0\farcs05.  We therefore calculate the detectability of
  exozodiacal emission by assuming this is equivalent to the
  detectability of a ring at the optimal radius.  Taking
  the optical depth of the zodiacal cloud, $\tau=5\times10^{-8}$
  \citep{dermott02astIII}, and assuming a ring of width $dr=0.5r$
  gives a fractional luminosity of
  $f=0.5(dr/r)\tau=1.25\times10^{-8}$.   If we consider a sunlike star
  at 10pc, the optimal detection radius would be centered at
  $r=0.5$AU.  Adopting the blackbody temperature for this dust
  ($T=278.3/\sqrt{r}$, see section 4.1) then the observed flux from
  exo-zodiacal dust should be 10$\mu$Jy at 10$\mu$m (using equation 6
  of \citealt{wyattreview}:
  $F_\nu=2.95\times10^{-10}B_{\nu}(\lambda,T)fr^2/d^2$ where $B_\nu$
  is the Planck function, $\lambda$ is the wavelength of observation
  assumed here to be 10$\mu$m and $d$ is the distance to the star here
  assumed to be 10pc) with higher flux for closer stars.  Thus we
  could expect to resolve discs down to $\lesssim$1 zodi out to
  10pc. This result is consistent with the previous expectations of
  the performance of METIS (as discussed in section 2.3.3 of
  \citealt{brandl}, METIS is expected to be able to resolve the
  exozodiacal emission in the 1AU region around stars at $<$10pc.) We
will also be able to resolve details of the strucutre of the few
bright discs already known at 10$\mu$m,  such as those around HD69830
and $\eta$ Corvi (K0V and F2V) which are believed to have dust in the
terrestrial planet regions ($\sim$1AU, \citealt{smithMIDI}).  
%

\section{Discussion}
\label{s:dis}

This sample contains 7 sources with excess infrared emission confirmed
either in this paper or with Spitzer data \citep{rieke, chen06}, and a
further 3 sources with excess emission in IRAS requiring confirmation.  The
SED fitting indicates that these objects are surrounded by dust at a
distance of between 2-60 AU (or alternatively two temperatures of dust
at 2 and 61AU for HD71155 and 4 and 24 AU for $\eta$ Tel).  These
regions are those in which we might expect the formation of giant
planets, and so it is important consider how the existence of this
dust emitting in the mid-infrared adds to our current understanding of
dust distributions in circumstellar regions.

\begin{figure}
\includegraphics[width=8cm]{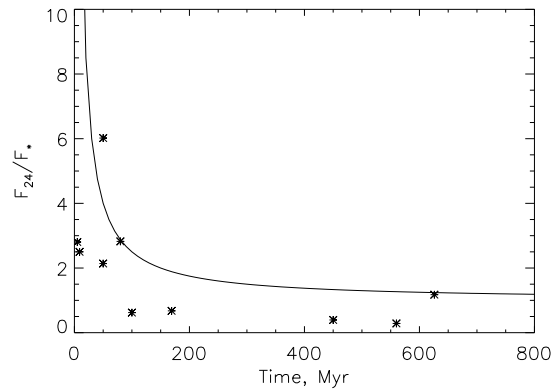}
\caption{\label{riekecomp} Plot of age versus fractional excess from
  SED fitting at 24 $\mu$m.  The line shows a $t^{-1}$ dependence,
  taken from \citet{rieke}, and this plot can be directly
  compared to their Figure 2.}
\end{figure}

\citet{rieke} looked at a sample of 266 A-type stars between 5-626
Myr old with MIPS at 24 and 70$\mu$m and examined the relationship
between fractional excess and the age of the central star.  They found
that the upper limit of excess emission generally fell off as
$time^{-1}$ for the stars with detected excess. Assuming the fits to
the SED profiles presented in this paper and plotting the predicted
24$\mu$m excess emission compared to stellar flux versus
age it is clear that the results presented here are in-line with this 
relationship (Figure \ref{riekecomp}).  Combining this with the age
spread in sources suggests we have a representative sample of A star
debris discs. 

HD71155 and $\eta$ Tel, the two systems with resolved dust
populations, have both been found to have multiple disc components. 
\citet{chen09} summarised the debris systems which have, through SED
fitting or resolved imaging, been identified as multiple component
discs.  Adding HD71155 to this sample brings the total number to 12
known ``Solar System analogues'' (as defined in
  \citealt{chen09}, systems with multiple
  component discs).  The FEPS survey on Spitzer surveyed
328 solar-like stars finding $\sim$10\% have 70$\mu$m emission
indicating the presence of cold debris, and of these 1/3 have SEDs
that are best fit by multiple temperature excesses
\citep{hillenbrand}.    From our sample of 10 debris disc sources (3
of which, $\lambda$ Gem, HD23281 and HD141795,  still require
photometric confirmation), 2 are multi-component discs (20\%;
consistent with the 1/3 rate given our small sample size).  These
results are further evidence that multiple component or extended discs
are common.  The traditional view of debris discs is that of a ring of
planetesimals residing outside any planetary system producing cold
dust, analogous to the Edgeworth-Kuiper belt (EKB) in the Solar System.
Multiple component discs could be seen as analogous to the Solar
System, although of course the
extrasolar discs detected to date are much brighter than the asteroid
belt and EKB ($L_{\rm{IR}}/L_\star \ge 10^{-4}$ for detected
extrasolar discs;  asteroid and Edgeworth-Kuiper belts
$L_{\rm{IR}}/L_\star = 10^{-8}-10^{-7}$ and $10^{-7}-10^{-6}$
respectively, \citealt{dermott02,stern96aa}). 

For sources with large amounts of emission from close to the star, the
origin of the dust is uncertain.  \citet{wyattsmith06} presented a
model based on the collisional evolution models of \citet{dd03} that
predicted a maximum dust luminosity dependent on age for a disc at a
given radius.  This maximum brightness occurs when discs just reach
collisional equilibrium, in which the size distribution of the
planetesimals in a disc is fixed as mass is transferred down through a
cascade of collisions to smaller and smaller sizes until they are
removed by radiation pressure.  Very massive discs process their mass
very quickly and are therefore short-lived, whereas sparse discs take
a long time to reach collisional equilibrium but are not very
luminous.  This model has been shown to accurately recreate the A star
debris disc population observed with Spitzer \citep{rieke} under
assumptions of a distribution of initial disc mass, radius, and current
age \citep{wyattsmith07}.  The predicted maximum luminosity for each
disc (or disc component) in this study is given in Table
\ref{results} ($f_{\rm{max}}=
  1.2\times10^{-6}r^{7/3}t_{\rm{age}}^{-1}$ from equation 20 in
  \citealt{wyattsmith07} where $r$ is the radius of the dust belt from
  our fits and $t_{\rm{age}}$ is the age of the system as given in
  Table \ref{sample}) .
Uncertainties in the model parameters
mean that only if $f_{\rm{IR}}/f_{\rm{max}} > 1000 $ do we take
this as evidence of
transient emission (level of excess cannot be produced by a
collisional cascade).  Most of our targets have
$f_{\rm{IR}}/f_{\rm{max}} < 1000$, with $\lambda$ Gem and the inner
disc of HD71155 exceeding this value, and HD141795 having
$f_{\rm{IR}}/f_{\rm{max}} \sim 550$.  Of these possibly transient
sources only HD71155 has confirmed excess emission (see Section 4.2),
and confirmation of the other discs would be required before
speculating on their origin.  

At an offset of 2AU around HD71155 the collisional lifetime of bodies
is short and thus we interpret the emission as evidence for a
transient dust producing event.  A recent massive collision in an
otherwise quiescent disc could produce a short lived increase in
excess, although as such events are likely to be rare (and the
resulting dust grains have short lifetimes) the probability of
witnessing such an event is low (see discussion in
\citealt{wyattsmith06}).  At an age of 169Myr (Table \ref{sample}),
ongoing terrestrial planet formation could be responsible for the
emission \citep{kenyon04}, with collisions between planetary embryos
resulting in large amounts of dust production.  The cooler belt at
61AU could represent a parent population of the hot dust emission as
well as producing spatially coincident dust dominating the excess at
longer wavelengths (a similar possibility exists for the older
Sun-like star  $\eta$ Corvi, \citealt{smithhot, smithMIDI}).  However,
the transport mechanism to get the dust to 2AU is unknown.  A
dynamical instability like that thought to have caused the Late Heavy
Bombardment (LHB) in the Solar System \citep[triggered by the
  migration of Jupiter/Saturn, see e.g.][]{gomes,
  levison}, during which large amounts of debris from the EKB
was thrown into the inner Solar System, could be
responsible. \citet{booth} concluded that such an event occured around
at most 12\% of Sun-like stars.  Around A stars, where the
70$\mu$m excess has been shown to exhibit a fall-off proportional to
time like the 24$\mu$m emission \citep[albeit with a longer decay
  time, ][]{su}, the statistics are less clear.  Resolving
the outer disc component would allow further examination of a possible
link between the two populations (for example, emission
spread inwards towards the hot dust population rather than confined to
a narrow could be evidence for a link).  A further possibility is that
the emission arises from dust grains not produced in collisions but in
the sublimation of a population of comets or one Super-comet.  These
possibilities were explored as the origin of the hot dust population
around HD69830 by \citet{beichman05}, who concluded that the
continuous generation of small grains by a population of comets would
require too large a mass reservoir to be the likely origin of the
dust. A single massive comet (Sedna-sized in the case of HD69830)
could release small dust grains over a few Myr if captured into a
close orbit \citet{beichman05}.  This mechanism could be responsible
for the 24$\mu$m dust population in more systems.

Dust emission that is from a transient event will necessarily show
temporal variation.  The difference between the IRAS photometry on
HD3003 and the measurements taken with MIPS and those presented in this
paper (see section 4.1) could be evidence of such evolution. If real,
this variance could be a reflection of the binary nature of the
system, with the orientation of the secondary as it proceeds on its
orbit changing the overall illumination of the system.  A
determination of the orbit of the binary will allow this possibility
to be checked.  Alternatively, if the temporal variance reflects
changing levels of dust or changing dust distributions then a
transient origin is more likely.  Taking the assumed circumstellar
radius of 4AU for the dust (see section 4.1) 
$f_{\rm{IR}}/f_{\rm{max}} = 379$, a high level but one at which we
would not state the emission must be transient conclusively (see above
and detailed discussion of uncertainties in
\citealt{wyattsmith06}).   However, if the dust is of transient
origin, the dust location needs not be stable, and the constraint of
$>$14.4AU for the binary semi-major axis need not hold.  In this
situation the determination of the orbit of the binary could
again greatly improve our understanding of this system.  Alternatively
resolving the dust distribution (as would be possible with METIS on
the E-ELT, see Figure \ref{futurepred}) could also inform our models
of the stability of the system, and thus the likely origin of the dust.  

Of the sources considered in this paper only $\eta$ Tel and
  HD71155 have had the location of the dust population confirmed by
resolved imaging.  For the remaining discs constraints have been
placed on the dust location, but degeneracies in the SED fitting in
particular (summarised in Section 5) mean that resolved imaging is
required to determine the true dust distributions in these systems,
and so constrain models for the dust origin particularly where
transience is inferred.  The predictions in section 5 can be used to
target sources most likely to be resolved with currently available
instruments, but as shown in Figure \ref{futurepred} most currently
known disc targets will require the use of MIRI to detect faint levels
of extended emission or the high resolution of METIS on the E-ELT to
resolve emission on small scales. High resolution will also be
important for the detection of sub-structure in the discs which could
indicate the presence of planets which will be important for
ascertaining the nature of these systems and distinguishing between
models for the origin of the dust.  Evidence for planets in the dust
distribution of debris disc systems include: sharp disc edges (as seen
around Fomalhaut, \citealt{kalas05}); clumps (Vega, \citealt{wyatt03};
similar structure may be observable in the EKB because of Neptune's
resonant Plutino population if the disc were brighter); warps ($\beta$
Pic, \citealt{augereau}); and asymmetries (HR4796A,
\citealt{wyatt99}).


\section{Summary}
\label{s:conc}

In this paper we have presented new observations of 11 early-type stars
which have been proposed to be debris disc hosts based on their IRAS
photometry.  We have used TIMMI2, VISIR, Michelle and TReCS data to
confirm excess emission and/or place constraints on debris discs for
the observed sample.  Our results are:
\begin{itemize}
\item For HD 3003, HD80950, and $\eta$ Tel our photometry yields an
  independent confirmation  of excess emission around the target.
  Subsequent analysis of the HD3003 system indicates that if the dust
  lies in a stable region it must be circumstellar and the binary must
  orbit at a semi-major axis of $\ge$14.4AU (assuming blackbody
  grains).   
\item Our data on 5 targets allow us to place quantitative limits on
  the location and level of emission of any dust in the system. For
  HD71155 these limits allow us to determine that the disc must have
  multiple components.   
\item  We use simple disc models to determine the region of disc flux
  versus radius parameter space for which discs can be resolved with
  currently available 8m mid-IR instruments.  This  
  technique successfully identified $\eta$ Tel as a resolvable disc,
  which was confirmed with TReCS \citep{smitheta}. 
\item We predict the parameter space of resolvable discs that could be
  opened by future instruments MIRI on the JWST and METIS on the E-ELT.
  Spatially extended disc structures will be best observed with MIRI
  because of their lower levels of surface brightness, whereas discs
  close to their central star (within $\sim$0\farcs3), or those with
  structure on small spatial scales, will be prime targets for E-ELT
  imaging which would be able to detect emission below 1 zodi
    out to 10pc. 
\end{itemize}

\begin{acknowledgements}
RS is grateful for the support of a Royal Commission for the
Exhibition of 1851 Fellowship.  Based on observations made with ESO
      Telescopes at the La Silla and Paranal Observatories under
      programme IDs 71.C-0312, 72.C-0041 and 74.C-0700. Also based on
      observations obtained at the Gemini Observatory, which is
      operated by the Association of Universities for Research in
      Astronomy, Inc., under a cooperative agreement with the NSF on
      behalf of the Gemini partnership: the National Science
      Foundation (United States), the Particle Physics and Astronomy
      Research Council (United Kingdom), the National Research Council
      (Canada), CONICYT (Chile), the Australian Research Council
      (Australia), CNPq (Brazil) and CONICET (Argentina).  
\end{acknowledgements}

\bibliographystyle{aa}  
\bibliography{/home/rachel/Work/PAPERs/thesis} 

\begin{appendix}
\label{s:app}

\section{Additional sources}

\subsection{Observed sources with no new limits} 

\begin{figure*}
\begin{minipage}{8cm}
\includegraphics[width=8cm]{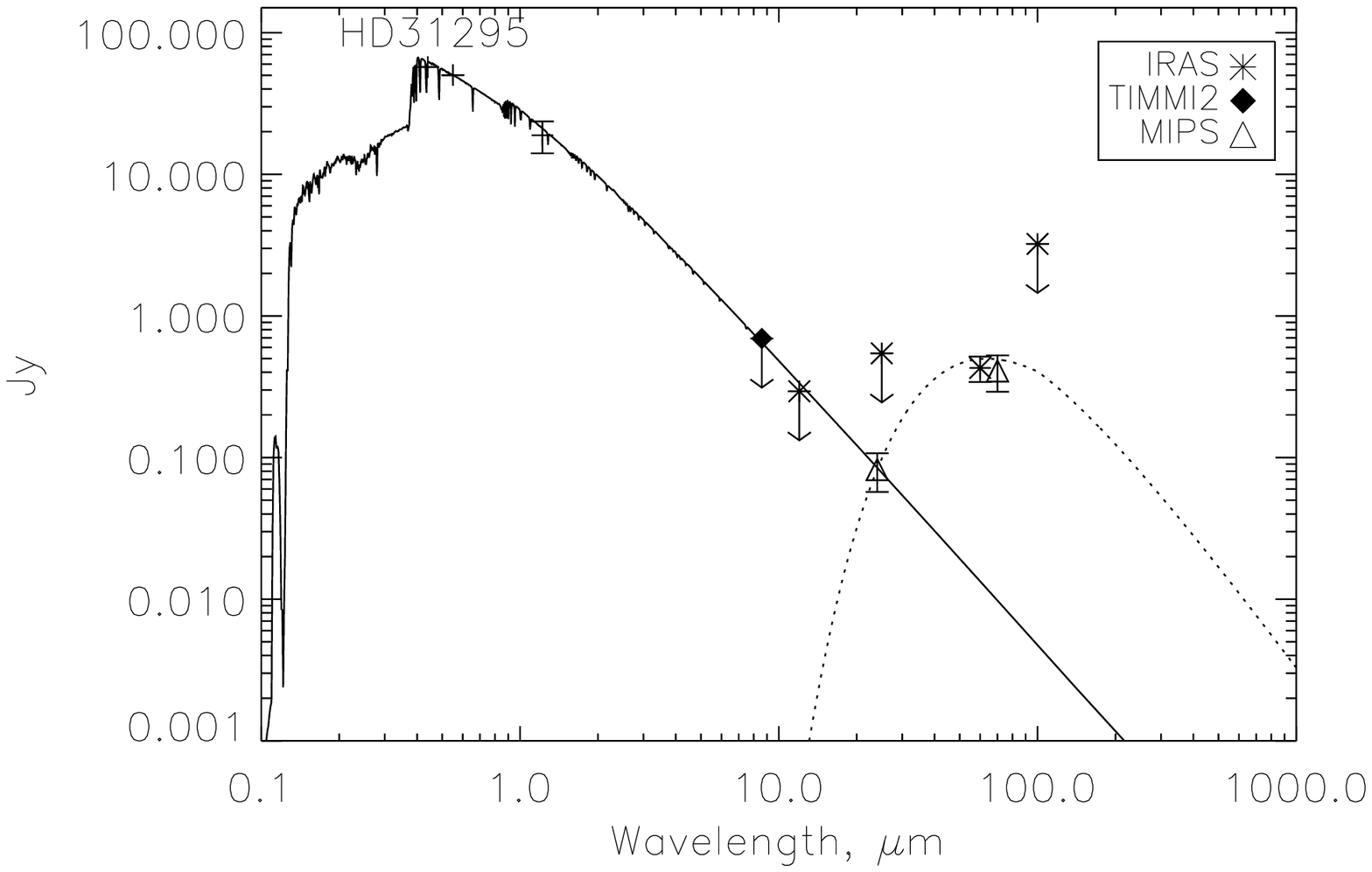}
\end{minipage}
\hspace{1cm}
\begin{minipage}{8cm}
\includegraphics[width=8cm]{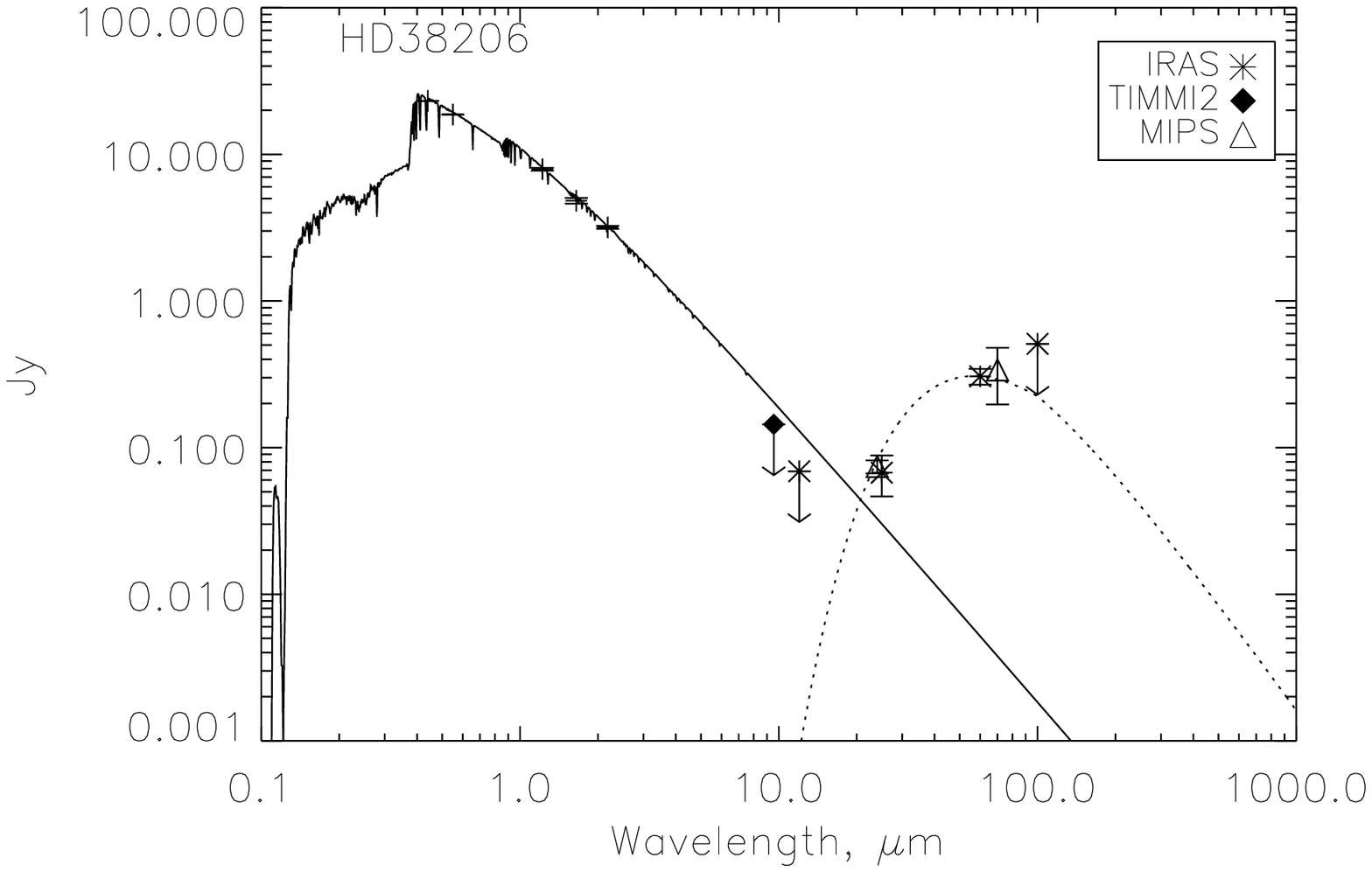}
\end{minipage}
\caption{\label{unconf} The SED fits of the excess emission of two
  stars with excess at longer wavelengths than that
  of the observations presented here. Left: HD 31295 with excess fit by
  a blackbody at 80K.  Right HD 38206 with excess fit by a blackbody at
  90K. }
\end{figure*}

\emph{HD 31295:} HD 31295 was first identified by \citet{sadakane} in their
sample of Vega-excess stars as a star with infrared excess identified
in the IRAS PSC (significant at 25$\mu$m, see Table
\ref{sample}). TIMMI2 N band observations do not confirm the excess
(Table \ref{observations}).  Results of Spitzer observations
\citep{jura04, su} place a
limit of $\leq$70mJy excess at 8.5$\mu$m, and the IRS spectrum shows
no evidence of excess at $\leq$20 $\mu$m. Further results
presented in \citet{su} showed that in MIPS 24 and 70$\mu$m photometry the
excess was confirmed. We find no evidence of background/companion
sources in the TIMMI2 field-of-view (such objects limited to
$<$33mJy). 

We fit the confirmed excesses with a blackbody at 80K (similar to
the Chen et al. fit of 90K, our errors are $\pm$12K from errors on the
excess detections), which translates to a radial offset of
52.6AU (1\farcs4, see Figure \ref{unconf} for SED). The image shows no
evidence for extension.  The disc flux expected in the observed TIMMI2
filter from the SED fit is $<$1mJy; longer wavelength observations
would be required to resolve/limit the disc extension around this
source.  \citet{martinez-galarza}
explored the possibility that this $\lambda$ Bootis star could 
be interacting with the ISM, and found that such
interaction could produce the observed excess.  With the current data
it is not possible to distinguish between
this theory and that of a circumstellar disc, although the fact that
this star lies within the local bubble (at 37pc, see Table \ref{sample})
means that the probability of the star lying within a cloud which
could produce the observed emission is low \citep{martinez-galarza}. 

\emph{HD 38206:} HD 38206 was first identified as a host of
mid-infrared excess by \citet{mannings} in their analysis of the IRAS
catalogues.  The TIMMI2 photometry presented here does not confirm the
excess (Table \ref{observations}).  Recent MIPS observations of this star 
\citep{rieke} have confirmed the 24 $\mu$m excess, with flux of 115
$\pm$ 12 mJy (expected photospheric flux 33$\pm$3 mJy, good
agreement with 
the IRAS measurements, see Table \ref{sample}). \citet{su} list the
MIPS 24$\mu$m photometry as 107mJy at 24$\mu$m and 342mJy at 70$\mu$m
(errors of 1.58mJy and 12.87mJy respectively do not include
calibration errors which are less than 5\% at 24$\mu$m and 10\% at
70$\mu$m). The excess emission is fitted with a blackbody at
90$\pm$10K 
(translating to a radial offset of  48.4 AU, 0\farcs70, see Figure
\ref{unconf} right and Table \ref{results}).  This fit suggests the
disc flux at the wavelength observed with TIMMI2 is $<$1mJy; as for HD
31295, only longer wavelength high resolution imaging will be able to
constrain or potentially resolve this disc's location and geometry. 

\subsection{Not a debris disc candidate}

\begin{figure}
\includegraphics[width=8cm]{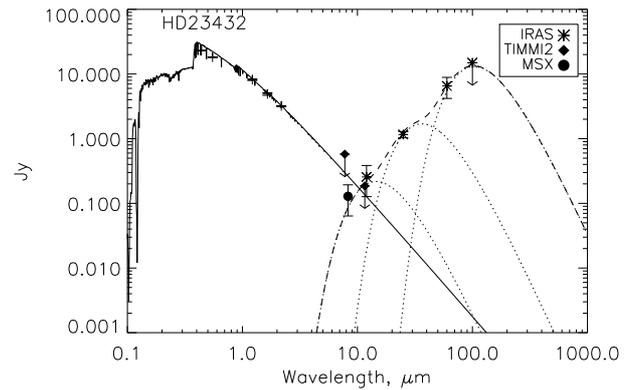}
\caption{\label{plot23432} The SED of HD 23432 with symbols at $>5\mu$m
  showing emission after subtraction of photospheric contribution.
  The excess is most likely due to the reflection nebulosity in this
  region. The TIMMI2 data points are highly untrustworthy due to the
  poor photometric conditions.  Blackbody fits for illustration
  purposes are at 200K, 80K and 28K.}
\end{figure}

\emph{HD 23432:} HD 23432 (asterope) was identified by \citet{oudmaijer}
as being amongst a sample of SAO stars with IRAS infrared
excess.  This star has an excess of 256 $\pm$ 43 mJy at 12 $\mu$m and
1159 $\pm$ 42 mJy at 25 $\mu$m (after subtraction of the
photosphere). In addition it also has excess at longer wavelengths:
6533 $\pm$ 780 mJy at 60 $\mu$m; and 13078 $\pm$ 4979 mJy at 100$\mu$m.  

The excess emission is not confirmed in the TIMMI2 observations of
HD 23432, as a flux of 141 $\pm$ 62 mJy at 11.6 $\mu$m is found compared
to an expected stellar flux of 136$\pm$7 mJy from a Kurucz profile fit (see
Table \ref{observations}).  
The TIMMI2 data points plotted on the SED (Figure \ref{plot23432}) are shown
with the calibration limits taken from the standards immediately
before and after the science observation.  The overall photometric
errors are much higher, with a change of calibration factor
over the course of the night of around 30\%.  Optical observations of
this Pleiades member show it to lie close to a diffuse reflection
nebula Ced 19h \cite{cederblad}. The shape of the
excess emission spectrum (fit here with blackbody emission at
  200$\pm$30K, 80$\pm$25K, and 28$\pm$3K, where errors on the
  temperatures arise from fitting the excess emission within
  3$\sigma$) suggests that this is the result of the emission 
from the reflection nebula and not from a dust population centred on
the star itself.  \citet{gorlova} found that the Spitzer 24$\mu$m
observations of this source were contaminated by flux from the
reflection nebula. Similar interactions with interstellar dust have been
shown to be responsible for excess emission towards other stars
\citep{kalas, gaspar}.

\end{appendix}

\end{document}